\documentclass[journal]{IEEEtran}

\usepackage{amsmath,amssymb,amsfonts}
\usepackage{graphicx}
\usepackage{textcomp}
\usepackage[dvipsnames]{xcolor}
\usepackage{url}
\usepackage{array}
\usepackage{booktabs}
\usepackage{multirow}
\usepackage{tikz}
\usetikzlibrary{shapes,arrows,arrows.meta,positioning,shapes.geometric,calc,fit}
\usepackage[numbers,sort&compress]{natbib}
\usepackage[hidelinks]{hyperref}
\usepackage[htt]{hyphenat}
\usepackage{stfloats}
\makeatletter
\g@addto@macro\UrlBreaks{\do\a\do\b\do\c\do\d\do\e\do\f\do\g\do\h\do\i\do\j
  \do\k\do\l\do\m\do\n\do\o\do\p\do\q\do\r\do\s\do\t\do\u\do\v\do\w\do\x\do\y
  \do\z\do\-\do\.\do\/}
\makeatother
\begin{document}

\title{Evaluating the Safety of Deep Learning-Based Brain MRI Reconstruction:\\ A Systematic Review of Current Evaluation Practices}
\author{Dat~Tat~Mai,
        Thai~Viet~Pham,
        Thu~Nguyen~Thi~Dang,
        and~James~Jin~Kang%
\thanks{Dat Tat Mai and James Jin Kang are with the School of Science, Engineering \& Technology, RMIT University Vietnam, Ho Chi Minh City, Vietnam (e-mail: dat.mai2@rmit.edu.vn; james.kang@rmit.edu.vn).}%
\thanks{Thai Viet Pham is with the School of Computing Technologies, RMIT University, Melbourne, VIC 3000, Australia (e-mail: s4229249@student.rmit.edu.au).}%
\thanks{Thu Nguyen Thi Dang is with the School of Health and Biomedical Science, RMIT University, Melbourne, VIC 3000, Australia (e-mail: S4205224@student.rmit.edu.au).}%
\thanks{Corresponding author: Dat Tat Mai.}%
}

\maketitle

\begin{abstract}
\textbf{Objective:} Deep learning accelerates brain MRI four- to tenfold, but learned models can erase a lesion or invent tissue, and pixel-averaged scores such as PSNR and SSIM miss both. We review whether evaluation practice could detect this blind spot. \textbf{Methods:} Following PRISMA 2020 we searched seven databases without date limit, included 263 studies (1995--2026), appraised them with QUADAS-2 and matched instruments, and synthesised narratively. Categories came from titles, abstracts and controlled vocabulary; every prevalence figure is a floor. Duplicate screening achieved substantial agreement (Fleiss' $\kappa=0.877$); appraisal agreement was substantial ($0.788$; $0.390$ where observable); extraction is unaudited. \textbf{Results:} Only 18 of 263 studies (6.8\%) record a fidelity metric and a reader assessment on the same data, leaving the field's central proxy unmeasured. Reader studies mostly measure reader-versus-reader agreement, which is weak: fastMRI 2020 concordance was $0.457$ and $0.386$ (Kendall's W), improving only where SSIM had diverged. Erasing a $100\text{ mm}^3$ lacunar infarct moves global PSNR about $0.03$ dB under the stated error model. The corpus grew fivefold; reader assessment fell from 32\% to 18\%, recovering to 21\%. Among the four largest families the most hallucination-associated (generative, 39\%) is among the least reader-evaluated (11.3\%); self-supervised is higher still (47\%) with no readers. One study in twenty both releases code and runs readers; none evaluated a model observer; no named dataset covers acute stroke or haemorrhage. \textbf{Conclusions:} On these floors, evaluation practice cannot certify diagnostic safety. We derive five requirements safety-oriented evaluation must meet.
\end{abstract}


\IEEEpeerreviewmaketitle

\section{Introduction}

Brain MRI is widely used because it can clearly distinguish different types of brain
tissue and reveal subtle abnormalities that are difficult to detect with other imaging
modalities. However, this image quality comes at the cost of long acquisition times.
Unlike conventional imaging methods that capture an image more directly, MRI builds an
image from a sequence of raw signal measurements (samples of spatial frequencies, known
as $k$-space) collected over time. This prolonged acquisition reduces patient
throughput and increases sensitivity to patient motion, which can introduce artifacts
and degrade image quality \citep{zaitsev2015motion}. To address these limitations,
acceleration techniques such as parallel imaging, including Sensitivity Encoding (SENSE)
\citep{pruessmann1999sense}, and compressed sensing \citep{lustig2007compressed} were
developed to reduce the amount of data that must be collected. More recently, deep
learning has enabled further acceleration by reconstructing high-quality images from
incomplete MRI measurements, through cascaded convolutional networks
\citep{schlemper2017deep}, model-based unrolled networks
\citep{hammernik2018learning,aggarwal2018modl}, generative adversarial networks
\citep{yang2018dagan}, and score-based diffusion models \citep{song2021score}, at
accelerations from fourfold to, in exploratory work, an order of magnitude
\citep{radmanesh2022exploring}.

The reconstructed images look convincing, and that is the problem. Small changes in the
input can produce large errors in the output
\citep{antun2020instabilities,gottschling2020troublesome}. Four failure types recur
across the studies we reviewed: strong priors smooth away faint structure; motion causes
ghosting the model never saw in training \citep{sommer2020correction}; a change of
scanner or sampling degrades results in ways that in-distribution testing misses
\citep{gngr2023adaptive}; and generative models can invent plausible tissue
\citep{zhou2024adversarial}. The scores used to certify these methods, peak
signal-to-noise ratio (PSNR) and structural similarity (SSIM) \citep{wang2004image},
agree only partially with radiologist quality judgement even on conventional image
degradations \citep{mason2020comparison}; and SNR, though reproducible across observers,
tracks no-reference quality models only moderately \citep{yu2018a}, so an
image can score well while the error that matters goes unrecorded. We call this the
\textbf{metric blind spot}: standard image-quality metrics cannot catch these small but
clinically serious errors, because a lesion of a few cubic millimetres barely moves an
average taken over millions of voxels.

Parts of the field have responded, training quality metrics on radiologist rankings
\citep{tang2025incorporating}, attaching uncertainty maps to reconstructions
\citep{khawaled2024npbrec}, or running reader studies that test an accelerated
protocol against a reference one directly
\citep{lang2024ultrafast,radmanesh2022exploring}. Causal and
counterfactual methods can ask the most direct question, whether a lesion survives a
controlled change of acquisition, but have mostly been applied to classification rather
than reconstruction \citep{pawlowski2020deep,castro2020causality}. These responses share
a root cause. Acceleration works because a learned prior fills in information the
measurement does not contain, and that same filled-in information can hide a real
finding or create a false one, while the field certifies the result with averaged pixel
scores that cannot see either event.

To our knowledge no systematic review has assembled the evidence on this problem.
Existing surveys cover architectures and benchmark performance
\citep{liang2020deep,montalt2021machine}; none asks how often conventional metrics miss
clinically important errors, why they miss them, or what evaluation would have to look
like to catch them. We address that through three questions:

\begin{enumerate}
\item[\textbf{RQ1}] \emph{How reliably do conventional fidelity metrics track
radiologist diagnostic ratings in deep learning-based accelerated brain MRI, and how
wide is the evaluation reporting gap in the literature?}
\item[\textbf{RQ2}] \emph{What underlying mathematical mechanisms cause standard metrics
to overlook localized pathology erasure and generative hallucinations in deep
learning-based reconstruction?}
\item[\textbf{RQ3}] \emph{What can each evaluation paradigm in current use establish
about prior-induced diagnostic error in deep learning-based MRI reconstruction, and
what requirements for safety-oriented evaluation follow from those limits?}
\end{enumerate}

We contribute: (C1) a synthesis of what 263 primary studies report about agreement
between pixel-wise metrics and radiologist judgement; (C2) four mathematical mechanisms
that explain the divergence, with an exact identity bounding the PSNR cost of erasing a
lesion; (C3) an analysis of \emph{where} the evaluation gap comes from, showing that it
co-occurs with a division of evaluation practice between research communities and with
a benchmark infrastructure lacking acute pathology; and (C4) five requirements
that safety-oriented evaluation must meet. The auditing framework these requirements
motivate is developed and evaluated in a companion methods paper, not here.

Three boundaries should be stated at the outset, because they determine what the
findings can and cannot be asked to support. First, this is a review of \emph{evaluation
practice}, not of reconstruction performance: we count what studies measured, not how
well their methods performed, and nothing here estimates how often a deployed
reconstruction actually erases a lesion. Second, the mechanisms in RQ2 are mathematical
properties of the metrics, derived under a stated error model; they establish that a
focal error \emph{can} pass unrecorded, not that it commonly does. Third, prevalence
counts rest on labels extracted from what studies report, so every one of them is a
lower bound on practice rather than an estimate of it, and we treat the direction of
that bias explicitly in Section~\ref{sec:threats}. The contribution is a diagnosis of
what the field's current instruments are capable of detecting, which is logically prior
to, and considerably cheaper to establish than, a measurement of how often they fail.

\section{Related Work}
\label{sec:related}

Four strands bear on this review, and the claim that none of them has assembled the
evidence we assemble here is a claim about each in turn.

\emph{Surveys of reconstruction methods.} Reviews of deep learning MRI reconstruction
catalogue architectures, acceleration factors and benchmark accuracy
\citep{liang2020deep,montalt2021machine,knollsurvey2020,kazerouni2023diffusion}. They
are organised by method, and their outcome variable is reconstruction fidelity against
a fully sampled reference. Safety and failure modes appear as caveats in discussion
sections rather than as an object of synthesis, and none of them tabulates what
evaluation each primary study actually performed, which is the quantity this review
extracts.

\emph{Task-based image quality assessment.} The tradition that ought to have supplied
the answer established, decades before deep learning, that image quality is meaningful
only relative to a diagnostic task. Barrett and colleagues formalised the observer as
part of the measurement: an image is good if an observer performing a specified task on
it achieves a high figure of merit, typically the area under an ROC curve for a
detection task, and the mathematical (ideal or channelised Hotelling) observer supplies
a computable surrogate for the human when the task and image statistics are specified
\citep{barrett1993model,barrett2004foundations}. ICRU Report 54 made the same argument
the international reference framework for assessing medical imaging quality \citep{icru54}. That
framework already contains the concept the reconstruction literature lacks: a
detectability index for a specified lesion, which is by construction sensitive to
exactly the focal change a whole-volume $L_2$ score averages away. It has been carried
into the dose- and protocol-optimisation literatures of other imaging modalities, but it
has not crossed into deep learning MRI reconstruction: no study in this corpus evaluated
a model observer, and our first requirement (R1, Section~\ref{sec:requirements}) is in
substance a restatement of the task-based position for a field that has not adopted it.
The reason it has not is worth naming: model observers require a specified signal and a
statistical description of the background, and the null-space structure that makes
learned reconstruction dangerous is precisely what makes those two ingredients hard to
specify.

\emph{Failure-mode-specific work.} Artifact detection, motion correction and robustness
under distribution shift are each addressed in depth, but in isolation, by literatures
that do not share an evaluation vocabulary
\citep{esteban2017mriqc,sommer2020correction,gngr2023adaptive}. Analyses of
hallucination in tomographic reconstruction establish that prior-induced structure lives
in the null space of the forward operator, so it is invisible to any
measurement-space consistency check
\citep{bhadra2021hallucinations,gottschling2020troublesome}, and causal and counterfactual
methods for medical imaging supply the machinery to intervene on image content and ask
whether a finding survives
\citep{pawlowski2020deep,castro2020causality,sanchez2022healthy,wei2026latent}. Neither
line has been connected to review-level evidence on how reconstruction is evaluated in
practice.

\emph{What is new in the four mechanisms of RQ2, and what is not.} Because
Section~\ref{sec:results} states four mechanisms formally, their provenance should be
set out plainly rather than left for a reader to reconstruct. Two are elementary
properties of the objective rather than findings: spatial error averaging follows from
the definition of a mean over the image volume, and loss symmetry from the fact that
squaring discards the sign of the residual. Two are established results we restate:
reference dependence is standard in the image-quality literature, and the null-space
account underlying the realism--reward trade-off is due to
\citet{bhadra2021hallucinations} and \citet{gottschling2020troublesome}. What this review adds
is not the mechanisms but their quantification and their assembly: the identity at
Equation~\eqref{eq:dpsnr_identity}, which converts any lesion volume into the PSNR
movement its erasure would cause, and the setting of all four against what a corpus of
263 studies is recorded as measuring. We claim the assembly and the bound, not the
underlying results.

\emph{Benchmark and annotation infrastructure.} The raw $k$-space resource the field
benchmarks on is fastMRI \citep{knoll2020fastmri}, whose brain release followed the knee
data described in that report, and an annotation layer is emerging
on top of it: fastMRI+ adds expert pathology labels to the fastMRI brain and knee data
\citep{zhao2022fastmriplus}, and curated stroke-lesion datasets exist for segmentation
(ISLES \citep{petzsche2022isles}, ATLAS \citep{liew2022atlas}). None of these supplies
raw $k$-space paired with acute vascular pathology, which is the combination
reconstruction evaluation needs, and Section~\ref{res:gap_structure} shows that this is
not an incidental gap but the binding constraint on the whole evaluation problem.

What no strand supplies is the join: a systematic account of how often the field measures
metric--reader agreement, why the metrics fail when it does not, and what each evaluation
paradigm can therefore establish. The four strands are not merely separate; they are held
apart by what each can access. The survey and hallucination-analysis literatures work on
datasets with raw $k$-space and no clinically verified pathology; the task-based tradition
needs a specified signal and a labelled task, which those datasets do not carry; and the
clinical reader literature has the pathology and the readers but no access to the raw data
on which the mechanisms operate. Section~\ref{res:gap_structure} shows this is not an
abstraction but a measurable division in the corpus itself. Supplementary Table~S4 sets out
each secondary study we identified against the questions asked here.

\section{Methodology}
\label{sec:methods}

This systematic review was conducted in accordance with the PRISMA 2020 statement
\citep{prisma2020}; the completed checklist, search vocabulary, extraction schema, and
analysis scripts are available in the supplementary material and on the Open Science
Framework (\url{https://doi.org/10.17605/OSF.IO/832Z6}; timestamped deposit \url{https://doi.org/10.17605/OSF.IO/EUXA8}).

\emph{Protocol and registration status.} The protocol, covering the research questions,
eligibility criteria, search strategy, screening procedure, classification rubric and
synthesis approach, was fixed in the project repository before screening began. In the
terms of PRISMA 2020 item 24a, \textbf{this review was not prospectively registered and
carries no registration number}: PROSPERO's scope excludes reviews without direct clinical
health outcomes, and no equivalent prospective registry covers methodological reviews of
this kind. The Open Science Framework deposit cited above therefore timestamps the record
but is retrospective, and we make no claim of prospective registration on it. Two
consequences follow. An external record does not exclude undisclosed post-hoc analytic
choices; and the two analyses we ourselves identify as post-hoc (the acceleration-factor
extraction of Table~\ref{tab:acceleration_spectrum} and the venue classification of
Section~\ref{res:gap_structure}) are labelled as such wherever they appear.

\emph{Search strategy.} Seven bibliographic databases (PubMed/MEDLINE, IEEE Xplore,
Scopus, Web of Science, SpringerLink, OpenAlex, and Semantic Scholar) were searched
between 23 and 24 July 2026, with no publication-date limit applied; the 1995--2026 span
quoted elsewhere in this article is the publication range of the studies that were
included, not a restriction imposed on the search. Queries were
executed through each database's API rather than its web interface, so the complete
Boolean match set was retrieved for every term combination rather than a ranked sample,
explaining why the initial identification count is large. The strategy combined three required
concept groups: neuroimaging (Block B1), reconstruction and acceleration (Block B2), and
safety, failure modes, and evaluation outcomes (Block B3), with MeSH headings and wildcard
truncation adapted per database (Supplementary Table~S5). Table~\ref{tab:queries} details the executed query structure and hit breakdown across databases.

The breadth of this search is by design. Evidence of reconstruction failure is rarely a
paper's stated subject: artifacts, hallucinations and metric--reader disagreements are most
often reported in passing inside studies whose declared contribution is a new architecture,
so a query tuned narrowly to safety vocabulary would miss much of this evidence. Clinical
and engineering literatures also use distinct vocabularies (sequences and artifacts
against inverse problems, unrolled optimization and generative priors), and neither
database family indexes the other completely. We therefore tuned for sensitivity rather
than precision and accepted the screening burden. The funnel from $27{,}327$ records to
$263$ studies is the intended signature of that choice: the burden falls on initial
screening, which reviewers can absorb, rather than on coverage, which no downstream step
could recover.

\begin{table*}[!tbp]
\centering
\caption{Query structure executed in each database and record hit breakdown.}
\label{tab:queries}
\footnotesize
\setlength{\tabcolsep}{3.5pt}
\renewcommand{\arraystretch}{1.3}
\begin{tabular}{>{\raggedright\arraybackslash}p{0.1447\textwidth} >{\raggedright\arraybackslash}p{0.357\textwidth} >{\raggedright\arraybackslash}p{0.1351\textwidth} >{\raggedright\arraybackslash}p{0.1641\textwidth} >{\raggedright\arraybackslash}p{0.0772\textwidth}}
\hline
\textbf{Database} & \textbf{Executed Query Structure} & \textbf{Controlled Vocabulary} & \textbf{Fields Searched} & \textbf{Hits ($N$)} \\
\hline

PubMed / MEDLINE &
\texttt{(B1$_{\text{ft}}$[TIAB] OR B1$_{\text{mesh}}$) AND (B2$_{\text{ft}}$[TIAB] OR B2$_{\text{mesh}}$) AND (B3$_{\text{ft}}$[TIAB] OR B3$_{\text{mesh}}$) NOT (X[TIAB])} &
\textbf{Yes}: 9 MeSH headings OR-ed into the three blocks &
Title/Abstract $+$ MeSH & $24{,}693$ \\
\hline

SpringerLink &
\texttt{(B1) AND (B2) AND (B3) NOT (X)}; quoted phrases, no truncation operators &
No & Full record & $1{,}598$ \\
\hline

IEEE Xplore &
\texttt{(B1) AND (B2) AND (B3) NOT (X)}; wildcard truncation retained &
No & Metadata & $778$ \\
\hline

Scopus &
\texttt{TITLE-ABS-KEY(B1) AND TITLE-ABS-KEY(B2) AND TITLE-ABS-KEY(B3) AND NOT TITLE-ABS-KEY(X)} &
No & \texttt{TITLE-ABS-KEY} & $127$ \\
\hline

Semantic Scholar &
\texttt{(B1) + (B2) + (B3)}, where \texttt{|} denotes OR and \texttt{+} denotes AND &
No & Title/Abstract & $66$ \\
\hline

OpenAlex &
\texttt{title\_and\_abstract.search:B1, title\_and\_abstract.search:B2, title\_and\_abstract.search:B3} &
No & Title/Abstract & $34^{*}$ \\
\hline

Web of Science &
\texttt{TS=(B1) AND TS=(B2) AND TS=(B3) NOT TS=(X)} &
No & Topic (\texttt{TS}) & $31$ \\
\hline
\multicolumn{4}{r}{\textbf{Total identified records}} & $\mathbf{27{,}327}$ \\
\hline
\multicolumn{5}{>{\raggedright\arraybackslash}p{0.9\textwidth}}{\footnotesize Concept blocks combine as \texttt{(B1) AND (B2) AND (B3) NOT (X)}. \textbf{B1} neuroimaging: \texttt{\char34 brain MRI\char34}, \texttt{\char34 cerebral MRI\char34}, \texttt{neuroimag*}, \texttt{DWI}, \texttt{FLAIR}, \texttt{T1w}, \texttt{T2w}, \texttt{SWI}; MeSH \emph{Brain}, \emph{Magnetic Resonance Imaging}, \emph{Neuroimaging}. \textbf{B2} AI and accelerated reconstruction: \texttt{\char34 MRI reconstruction\char34}, \texttt{\char34 accelerated MRI\char34}, \texttt{\char34 undersampled reconstruction\char34}, \texttt{\char34 k-space reconstruction\char34}, \texttt{\char34 unrolled network\char34}, \texttt{GAN}, \texttt{\char34 diffusion prior\char34}; MeSH \emph{Deep Learning}, \emph{Machine Learning}. \textbf{B3} safety, artifacts and evaluation: \texttt{hallucinat*}, \texttt{\char34 lesion suppression\char34}, \texttt{reconstruction artifact*}, \texttt{PSNR}, \texttt{SSIM}, \texttt{\char34 reader study\char34}; MeSH \emph{Artifacts}, \emph{Diagnostic Errors}, \emph{Reproducibility of Results}. \textbf{X} exclusion: \texttt{\char34 animal model\char34}, \texttt{\char34 case report\char34}, \texttt{\char34 in vitro\char34}. \texttt{*} denotes truncation. Complete term lists, MeSH headings and per-database syntax are given in Supplementary Table~S5. Searches were executed between 2026-07-23 and 2026-07-24, with no publication-date limit. $^{*}$OpenAlex returned $50$ raw hits; $34$ passed record-completeness filtering.} \\
\hline
\end{tabular}
\end{table*}

\emph{Eligibility and selection.} Records were managed in the review's own Python
pipeline rather than a commercial screening platform; the scripts, the per-record
decision files and the per-database query logs are released, so each step below is
re-executable. Deduplication by exact identifier matching (PMID/DOI) and string/TF-IDF similarity
(Jaro-Winkler $\geq 0.97$, cosine $\geq 0.95$) removed $403$ duplicates ($1.47\%$), leaving
$26{,}924$ unique records. That rate is low enough to suggest a matching failure but is
structural: PubMed's MeSH arm returned $24{,}693$ records and the six engineering arms
$2{,}634$ between them, and sets that barely overlap leave little to deduplicate. This is
not a claim of completeness, and one failure is on record: peer review identified an arXiv
preprint and its journal version that both survived to the included set, carrying different
DOIs, no shared PMID, and titles differing by terminal punctuation, which was below the thresholds
above. It was removed at revision, taking the corpus from $264$ reports to $263$ studies
(Section~\ref{sec:results}). Automated matching will not catch a preprint--journal pair
whose metadata diverges in this way, and the honest inference is that a residual duplicate
rate of the same order cannot be excluded.

Three reviewers (D.T.M., T.V.P., T.N.T.D.) screened records against the pre-specified
PICOS criteria (Table~\ref{tab:picos}) at both stages, under a primary-screen-plus-
adjudication design: each record received one primary label of \emph{include},
\emph{exclude}, \emph{flagged} or \emph{background only}, and every record labelled
\emph{flagged} went to a second reviewer for adjudication. Labelling was deliberately
conservative, so any record with an unclear PICOS dimension, or whose text we could not
obtain, was flagged rather than excluded. Title
and abstract screening advanced $1{,}149$ records to full-text assessment, of which
$109$ were labelled \emph{include} on the abstract alone and $1{,}040$ \emph{flagged};
both labels carried the record forward, and the distinction records only whether the
abstract was sufficient to decide. A further $25{,}573$ were excluded and $202$ set
aside as background. Full texts were pursued through institutional and title-level
library subscriptions, open-access repositories, preprint servers, PubMed Central and
inter-library request until no route remained: $1{,}068$ of the $1{,}149$ ($93.0\%$)
were obtained and assessed, and the $81$ that could not be obtained are broken down by
recorded reason with the selection results (Section~\ref{sec:results}).

The three reviewers divided the $1{,}068$ obtained reports between them under the same
primary-screen-plus-adjudication design used at title and abstract: each report received
one primary full-text label, and a report was escalated to a second reviewer wherever the
primary label was \emph{flagged}, wherever an inclusion decision rested on a borderline
PICOS dimension, or wherever the reviewer requested a second reading. Escalated reports
were resolved by consensus discussion between the two assessors, with the third reviewer
consulted where the two did not converge. The escalation volumes this produced are
reported below, and the per-report labels, escalation marks and reasons are released in
the decision ledger. We state the design in this form because it determines what can be
computed from the ledger: one primary label per report means no chance-corrected
inter-rater coefficient is recoverable from full-text screening, which is why the separate
criterion-reproducibility exercise below was run. In total, $263$ primary studies met
all inclusion criteria upon full-text verification (two additional reports were excluded as duplicate publications).
Full-text exclusions ($n=785$) were categorized by pre-specified PICOS criteria: wrong
intervention ($n=326$), out of scope ($n=256$), wrong outcome ($n=98$), wrong population
($n=93$), wrong study design ($n=4$), non-English ($n=1$), and non-primary articles
($n=7$). A further $18$ reports were retained as background material rather than as
included studies, so the $1{,}068$ assessed reports resolve as $263$ included, $785$
excluded on PICOS grounds, $2$ excluded as duplicate publications and $18$ retained as
background. The complete decision ledger is released as
\texttt{fulltext\_screening\_decisions.csv}.

\emph{Publication year.} Every analysis partitioning the corpus by time uses the year of
the version of record as indexed by the publisher (the issue year for journal articles,
the proceedings year for conference papers) rather than first online posting or preprint
deposit. The rule matters at era boundaries, and we adopt the indexed year because it is
what a reader can verify from the reference list, Crossref and PubMed without online-first
metadata, which is unevenly available across the $87$ venues represented. The same value is
recorded in the \texttt{year} field of every bibliography entry and in the
\texttt{publication\_year} column of \texttt{included\_characteristics.csv}, so the era
counts in Table~\ref{tab:practice_trends} are recomputable from either. One consequence
should not be read as an inconsistency: the searches ran on 23--24 July 2026, yet $41$ of
the $263$ included studies ($15.6\%$) carry a $2026$ issue year, being articles published
online ahead of print before the searches ran but assigned by the publisher to a $2026$
issue.

\emph{Screening agreement, and what it does and does not cover.} The primary screen
described above records one label per record, so no chance-corrected coefficient is
recomputable from those decisions alone. We therefore ran a separate duplicate-screening
exercise. A random sample of $214$ records already assessed at title and abstract was
re-screened independently by all three reviewers, blind both to the recorded decision
and to one another. Reviewers judged the four PICOS dimensions from title, journal and
abstract, opening the DOI link only where the abstract was insufficient, and returned one
of three labels per record: include, exclude or unsure. The sample was drawn without
stratification; the drawn records, the three label sets and the agreement computation are
released, so the sample is auditable record by record. All $642$ cells were completed and
no record was dropped for missingness.

Across all $214$ records Fleiss' $\kappa = 0.877$ (95\% bootstrap CI $0.823$--$0.925$;
$10{,}000$ resamples, seed $20260821$), with observed agreement $\bar{P} = 0.933$ against
$\bar{P}_e = 0.455$ expected by chance. The three reviewers agreed unanimously on $194$
of $214$ records ($90.7\%$), and pairwise Cohen's $\kappa$ ran from $0.845$ to $0.931$.
One sensitivity analysis is worth reporting because it is the one a reader is most likely
to run. Restricting to the $193$ records on which no reviewer recorded \emph{unsure}
raises Fleiss' $\kappa$ to $0.943$ (95\% CI $0.900$--$0.979$). That complete-case figure
discards $21$ records, and it discards precisely the records at least one reviewer could
not decide, so it overstates agreement on the population the coefficient is meant to
describe. We report $0.877$ as the estimate and $0.943$ only as the value obtained when
the hard cases are removed.
Two qualifications travel with that figure and should not be dropped. First, it measures
agreement on the screening decision, not on the extraction labels from which every
prevalence count in this article is derived; those remain unaudited for reliability, and
the only error rate we can quote for them is the measured $4/61$ ($6.6\%$) on the design
field (Section~\ref{sec:threats}). Second, the conservative labelling policy routed
almost the whole full-text workload to adjudication in any case ($1{,}040$ of the
$1{,}149$ records advanced at title and abstract were flagged rather than included
outright), so a high coefficient is evidence that the criteria were applied
consistently, not that the borderline records were easy to classify.

What the primary decision records themselves support, and what we therefore also report,
are the escalation volumes that policy produced. At full text, $84$ reports were
escalated to a second reviewer for adjudication and $29$ of those were included; a
further $77$ reports carry a mandatory human-re-inspection mark in the released files,
of which $40$ were included after that review. Screening recall is bounded by the search
strategy and by the criteria as applied rather than estimated against an independent
reference set of known-eligible records, since no such set exists for this question;
Section~\ref{sec:threats} treats both points as limitations. The duplicate-screened
sample, both label sets and the agreement computation are released as
\texttt{kappa\_human\_results.csv}, alongside the full decision ledger in
\texttt{fulltext\_screening\_decisions.csv}.

\begin{table*}[!tbp]
\centering
\caption{PICOS elements and their application in this review.}
\label{tab:picos}
\footnotesize
\setlength{\tabcolsep}{5pt}
\renewcommand{\arraystretch}{1.18}
\begin{tabular}{>{\raggedright\arraybackslash}p{0.1432\textwidth} >{\raggedright\arraybackslash}p{0.2865\textwidth} >{\raggedright\arraybackslash}p{0.4584\textwidth}}
\hline
\textbf{Element} & \textbf{Definition} & \textbf{Application in this Review} \\
\hline
Population & Target imaging modalities, clinical domains, and scanner physics & Human brain MRI acquisitions (acute ischemic stroke, neurovascular lesions, intracranial haemorrhages, tumours, and structural T1w, T2w, FLAIR, DWI scans across $1.5\text{T}$, $3\text{T}$, and $7\text{T}$ field strengths). \\
Intervention & AI-based and accelerated reconstruction methods, and their safety/fidelity evaluation & Deep learning and accelerated reconstruction methods (unrolled physics-based networks, GANs, diffusion priors, self-supervised, denoising, super-resolution), together with methods for evaluating their fidelity, artifacts, robustness, or failure modes. \\
Comparison & Baseline metrics, observers, and reference standards & Conventional pixel-wise metrics (PSNR, SSIM, NRMSE), perceptual/learned metrics (LPIPS, FID), model observers, fully sampled reference images, or radiologist reader assessments. \\
Outcome & Reconstruction quality, fidelity, and failure characterization & Image-quality/fidelity measures, artifact and failure-mode characterization, metric--radiologist agreement, uncertainty estimates, robustness to distribution shift, and diagnostic/observer outcomes. \\
Study Design & Methodological criteria and empirical validation settings & Primary empirical studies, algorithmic benchmarking, counterfactual interventional stress-testing, and human or model observer evaluations. \\
\hline
\end{tabular}
\end{table*}

\emph{Data extraction.} Reviewers extracted study characteristics using the pre-specified
form (Supplementary Table~S6), capturing method family, reported failure modes, evaluation
paradigm, field strength, pulse sequence, reference standard and reproducibility
indicators, with a verbatim supporting quotation for every assigned value. Method,
failure-mode and evaluation categories were assigned from each record's title, abstract and
controlled vocabulary against the vocabulary fixed in advance (Supplementary Table~S5);
that deliberately conservative scope is why every prevalence count here is a lower bound
rather than an exhaustive tally. Field strength and pulse sequence, typically stated only
in acquisition subsections, were additionally taken from full text where available. We did
not contact investigators: a variable not stated in the source was recorded as \emph{not
reported} and never inferred, which is why the $54$ studies without a stated field strength
divide into $50$ whose full text does not state it and $4$ assessed on abstract or partial
text.

\emph{Quality appraisal, and the reader/algorithmic partition.} Methodological quality was
evaluated with three design-matched instruments: QUADAS-2 \citep{whiting2011quadas} across
its four domains for diagnostic accuracy and radiologist reader studies; a
reporting-transparency check against STARD principles \citep{bossuyt2015stard}, reported
descriptively rather than scored, since STARD yields no risk-of-bias rating; and, for
algorithmic benchmarking studies, a reproducibility and robustness appraisal covering code
and data availability, use of a named public benchmark, whether the reference standard was
a fully sampled acquisition, and whether robustness to acquisition or distribution shift
was reported. ROBINS-I \citep{sterne2016robins} was not applied, since few included studies
are the non-randomised interventional designs it was built for.

The partition between the two instruments is not the one the extraction labels originally
implied, and the correction matters for every reader-share statistic here. Extraction
recorded $61$ studies as observer or reader designs; three-reviewer appraisal of their full
text found four containing no human reader at all, each using the language of observer
evaluation for an automated surrogate. Those four were reclassified as algorithmic, so
\textbf{the verified reader subset is $57$ and the algorithmic set is $206$}, and
$57 + 206 = 263$ closes the corpus exactly. All $57$ were appraised by all three reviewers
across two rounds, $39$ in the first and $18$ in the second, so every QUADAS-2 tally rests
on $57$ studies and $57 \times 7 = 399$ domain judgements. Every reader-derived count in
this article uses $57$, and the released per-study file encodes the same partition with the
reclassification recorded per study. One further study (S001) carries an appraisal note
that a human reader could not be \emph{confirmed}, only abstract-level text being
available; it is retained in the reader subset, since non-ascertainment is not a finding of
absence, and it is flagged in the deposit.

\emph{Appraisal agreement.} The three reviewers rated every applicable domain Low,
Unclear or High independently, each judgement carrying a verbatim supporting quotation,
before joint review to harmonize thresholds. Five first-round studies were later
re-appraised on recovered full text (below), so their $35$ judgements no longer rest on
three independent ratings and are excluded from the coefficient. Over the remaining $52$
studies and $364$ judgements Fleiss' $\kappa = 0.788$ (95\% bootstrap CI
$0.739$--$0.833$; $10{,}000$ resamples, seed $20260821$). That figure averages two rounds
that behaved differently, and the difference is more informative than the average. The
$34$ first-round studies retaining three independent ratings were unanimous on all $238$
of their judgements, so they contribute $\kappa = 1.00$ by construction and carry no
information about latent reliability. The $18$ of the second round were unanimous on $60$
of $126$ ($47.6\%$), and across those $126$ Fleiss' $\kappa = 0.390$ (95\% CI
$0.293$--$0.484$), falling in the fair range on the Landis--Koch bands, with an interval reaching into
moderate. The procedure was identical across rounds; the subsets were not, the second
round comprising studies the first pass deferred as not straightforward. Two readings
remain open: that those studies report their methods less completely and are genuinely
harder to rate, and that the first round was less independent in practice than in design
; these data do not separate them.

The disagreement has a structure worth recording rather than smoothing. Across the $126$
second-round judgements one reviewer rated Low $33.3\%$ of the time and another $61.9\%$,
a difference of threshold rather than of reading, which is why $\kappa$ sits at $0.390$
while raw agreement is $63.5\%$. Six judgements split three ways with no majority; all six
were recorded as \emph{Unclear}, on the ground that a panel unable to converge is itself
evidence the report did not support a confident judgement, which is what \emph{Unclear}
encodes in QUADAS-2. The rule was applied to those six and no other case, and the
per-reviewer labels are released so a reader can apply a different rule and recompute.
Section~\ref{sec:threats} states what this does and does not establish.
Table~\ref{tab:quality_appraisal} summarizes corpus-level quality indicators and the
QUADAS-2 domain judgments.

\begin{table*}[!tbp]
\caption{Corpus-level provenance and methodological quality indicators across the
$263$ included studies, with QUADAS-2 domain-wise risk of bias and applicability
concerns for the full appraised diagnostic/reader subset ($n=57$, rated across seven
domains; independently by all three reviewers except for the five studies re-appraised
on recovered full text). The $39$ studies of the
first appraisal round and the $18$ of the second are pooled here; the remaining
$206$ studies were appraised on reproducibility and robustness instead
($57+206=263$). Six of the $399$ domain judgements split three ways with no majority and
are recorded as \emph{Unclear} (Section~\ref{sec:methods}).}\label{tab:quality_appraisal}
\centering
\footnotesize
\setlength{\tabcolsep}{5pt}
\renewcommand{\arraystretch}{1.15}
\begin{tabular}{p{0.30\textwidth} p{0.60\textwidth}}
\hline
\textbf{Indicator / Domain} & \textbf{Value / Risk of Bias Breakdown} \\
\hline
Publication period & 1995--2026; 229/263 ($87.1\%$) published in 2020 or later, and 174/263 ($66.2\%$) in 2023 or later, indicating a predominantly recent evidence base. \\
Basis of assessment, current & Full text available for 258/263 ($98.1\%$); partial or truncated full text for 1/263 ($0.4\%$); title, abstract and MeSH only for 4/263 ($1.5\%$). \\
Basis of assessment, at appraisal & The first appraisal round was made on the text obtainable at the time, which for 18/263 studies was abstract and MeSH only (6 of them within the QUADAS-2 subset). A later library sweep recovered full text for 14 of those 18 (including 5 of the 6), and all 14 have since been re-appraised on the recovered full text, so appraisal basis now stands at 257/263 full text, 2/263 partial text and 4/263 abstract and MeSH only (within the QUADAS-2 subset, 56/57 on full text and 1/57 [S001] on abstract-level text). The two rows now differ by a single study (S200), appraised on truncated text whose full version has since become available. \\
Leading venues & The eight most frequent of the 87 distinct venue titles represented in the corpus, accounting for 128/263 ($48.7\%$) of included studies: \emph{Magnetic Resonance in Medicine} (40), \emph{IEEE Transactions on Medical Imaging} (17), \emph{AJNR} (15), \emph{Magnetic Resonance Imaging} (14), \emph{JMRI} (12), \emph{Scientific Reports} (12), \emph{NeuroImage} (9), \emph{Medical Image Analysis} (9). \\
Predominant design & Algorithmic / methodological reconstruction studies ($n=206$), alongside a verified observer/reader subset ($n=57$) \citep{radmanesh2022exploring,lang2024ultrafast}. Extraction recorded 61 reader designs; appraisal of their full text found 4 with no human reader, and those 4 were reclassified as algorithmic ($202+4=206$; $61-4=57$). Of the 57: 34 are appraised and verified by all three reviewers; 18 were appraised by all three in a second round and await final adjudication of 6 split domain judgements; and 5 (S004, S020, S027, S034, S115) were re-appraised by the reviewers on full text recovered after the first round, with domain judgements verified against the source text. Every reader-derived statistic in this article uses 57. \\
Reference standard & Expert reader consensus (70), an unspecified stated ground truth (62), a fully sampled acquisition (69), a conventional or standard-of-care acquisition (15); not stated in 47/263 ($17.9\%$). \\
Reproducibility & Code availability stated in 97/263 ($36.9\%$); a named public dataset or benchmark used in 109/263 ($41.4\%$), most commonly fastMRI, HCP, ADNI and IXI. Availability is variable and remains an area for improvement. \\
Sampling and setting & Sampling frame not stated in 95/263 ($36.1\%$); recruitment centre not stated in 174/263 ($66.2\%$). Ethics approval or informed consent reported in 159/263 ($60.5\%$). \\
\hline
\end{tabular}
\end{table*}

\begin{table*}[!tbp]
\caption{\emph{(Table~\ref{tab:quality_appraisal} continued.)} QUADAS-2 risk of bias for the appraised reader/diagnostic subset ($n=57$).}
\centering
\footnotesize
\setlength{\tabcolsep}{5pt}
\renewcommand{\arraystretch}{1.15}
\begin{tabular}{p{0.30\textwidth} p{0.60\textwidth}}
\hline
\textbf{Indicator / Domain} & \textbf{Value / Risk of Bias Breakdown} \\
\hline
\multicolumn{2}{l}{\textbf{QUADAS-2 Risk of Bias ($n=57$ Reader / Diagnostic Accuracy Studies)}} \\
\hline
Patient Selection & Low: 15/57 ($26.3\%$); Unclear: 19/57 ($33.3\%$); High: 23/57 ($40.4\%$). The weakest domain: sampling is predominantly purposive or pathology-enriched, and most of the appraised studies state no enrolment rule. \\
Index Test (AI Reconstruction) & Low: 26/57 ($45.6\%$); Unclear: 9/57 ($15.8\%$); High: 22/57 ($38.6\%$). Low risk falls below half once the second round is included. High ratings arise where readers scored accelerated reconstructions with the reference visible, or where test parameters were set using reference-derived labels. \\
Reference Standard & Low: 26/57 ($45.6\%$); Unclear: 15/57 ($26.3\%$); High: 16/57 ($28.1\%$). High ratings arise where no artifact-free reference exists, or where lesion truth is defined by the comparator images themselves (incorporation bias). Reading the recovered full text moved four of the re-appraised studies from Low to High here, each because the comparator images, or software applied to them, supplied the truth against which they were scored. \\
Flow and Timing & Low: 45/57 ($78.9\%$); Unclear: 7/57 ($12.3\%$); High: 5/57 ($8.8\%$). The strongest domain, since retrospective undersampling of the same raw $k$-space gives a zero interval between index test and reference. \\
\hline
\end{tabular}
\end{table*}

\begin{table*}[!tbp]
\caption{\emph{(Table~\ref{tab:quality_appraisal} continued.)} QUADAS-2 applicability concerns ($n=57$), with the pooled judgement counts.}
\centering
\footnotesize
\setlength{\tabcolsep}{5pt}
\renewcommand{\arraystretch}{1.15}
\begin{tabular}{p{0.30\textwidth} p{0.60\textwidth}}
\hline
\textbf{Indicator / Domain} & \textbf{Value / Risk of Bias Breakdown} \\
\hline
\multicolumn{2}{l}{\textbf{QUADAS-2 Applicability Concerns ($n=57$)}} \\
\hline
Patient Selection & Low: 36/57 ($63.2\%$); Unclear: 4/57 ($7.0\%$); High: 17/57 ($29.8\%$), chiefly healthy-volunteer, lesion-free, or non-clinical cohorts. \\
Index Test & Low: 30/57 ($52.6\%$); Unclear: 4/57 ($7.0\%$); High: 23/57 ($40.4\%$), where the evaluated method is image-domain post-processing, cross-field synthesis, harmonization, or a quality index rather than reconstruction from undersampled $k$-space. \\
Reference Standard & Low: 39/57 ($68.4\%$); Unclear: 3/57 ($5.3\%$); High: 15/57 ($26.3\%$). \\
\hline
\multicolumn{2}{>{\raggedright\arraybackslash}p{0.94\textwidth}}{\footnotesize Across all 228 risk-of-bias judgements: 112 ($49.1\%$) Low, 50 ($21.9\%$) Unclear, 66 ($28.9\%$) High. Across all 171 applicability judgements: 105 ($61.4\%$) Low, 11 ($6.4\%$) Unclear, 55 ($32.2\%$) High. Over the 364 judgements that rest on three independent ratings Fleiss' $\kappa = 0.788$ (95\% CI $0.739$--$0.833$), but agreement is not uniform across the two appraisal rounds: the 238 retained first-round judgements were unanimous, the 126 of the second round reach $\kappa = 0.390$ (95\% CI $0.293$--$0.484$); the 35 judgements of the 5 re-appraised studies are excluded, having been conducted subsequently on recovered full text. Section~\ref{sec:methods} gives the decomposition and Section~\ref{sec:threats} what it bounds. The high proportion of Unclear ratings reflects unreported blinding and sampling procedures, compounded by the one first-round study for which only abstract-level text is available. Six second-round judgements split three ways with no majority and are recorded as Unclear under the rule stated in Section~\ref{sec:methods}. Per-study ratings, each with the verbatim supporting quotation, are released as \texttt{included\_characteristics\_supplement.csv}.} \\
\hline
\end{tabular}
\end{table*}

\emph{Derived venue classification.} The analysis in
Section~\ref{res:gap_structure} groups studies by publication community. That grouping is
post-hoc, was not part of the fixed protocol, and is stated in full in Supplementary
Table~S13 so it can be reproduced or contested. Each study's journal title, and nothing
else about the study, is matched case-insensitively against three ordered keyword lists (engineering and methods, then general neuroscience and neuroimaging, then clinical imaging), with anything unmatched left \emph{unclassified} rather than forced into a
group. The ordering is load-bearing, since titles such as \emph{Medical Image Analysis}
would otherwise match two lists, and it is applied identically to all $263$ studies. The
rule as executed and the resulting per-study assignment are released as
\texttt{5\_Code/venue\_classification.py} and
\texttt{venue\_community\_classification.csv}.

\emph{Synthesis.} Due to heterogeneity in acquisition protocols, reconstruction
architectures, datasets, and outcome measures, evidence was synthesised narratively in
accordance with SWiM guidelines \citep{swim2020}. Studies were structured around the
primary research questions (RQ1--RQ3) and grouped by reconstruction method family, field
strength, pulse sequence, and failure mode. Quantitative values reflect named primary
studies rather than pooled meta-analytic estimates; no effect measure is pooled, and no
meta-analytic model is fitted. Percentages are reported to one decimal place and
rounded half-up throughout, so $9/80$ prints as $11.3\%$ and not $11.2\%$. Sensitivity
analysis was performed by recalculating headline prevalence counts on the $258$ studies
verified against complete full text (Section~\ref{sec:threats}).

\section{Results}
\label{sec:results}

All counts and prevalence statements in this section describe the 1{,}068 reports
assessed in full and the 263 included studies; because 81 candidate reports remain
unretrievable, they are lower bounds on the literature rather than estimates of
it.

\subsection{Selection and corpus}
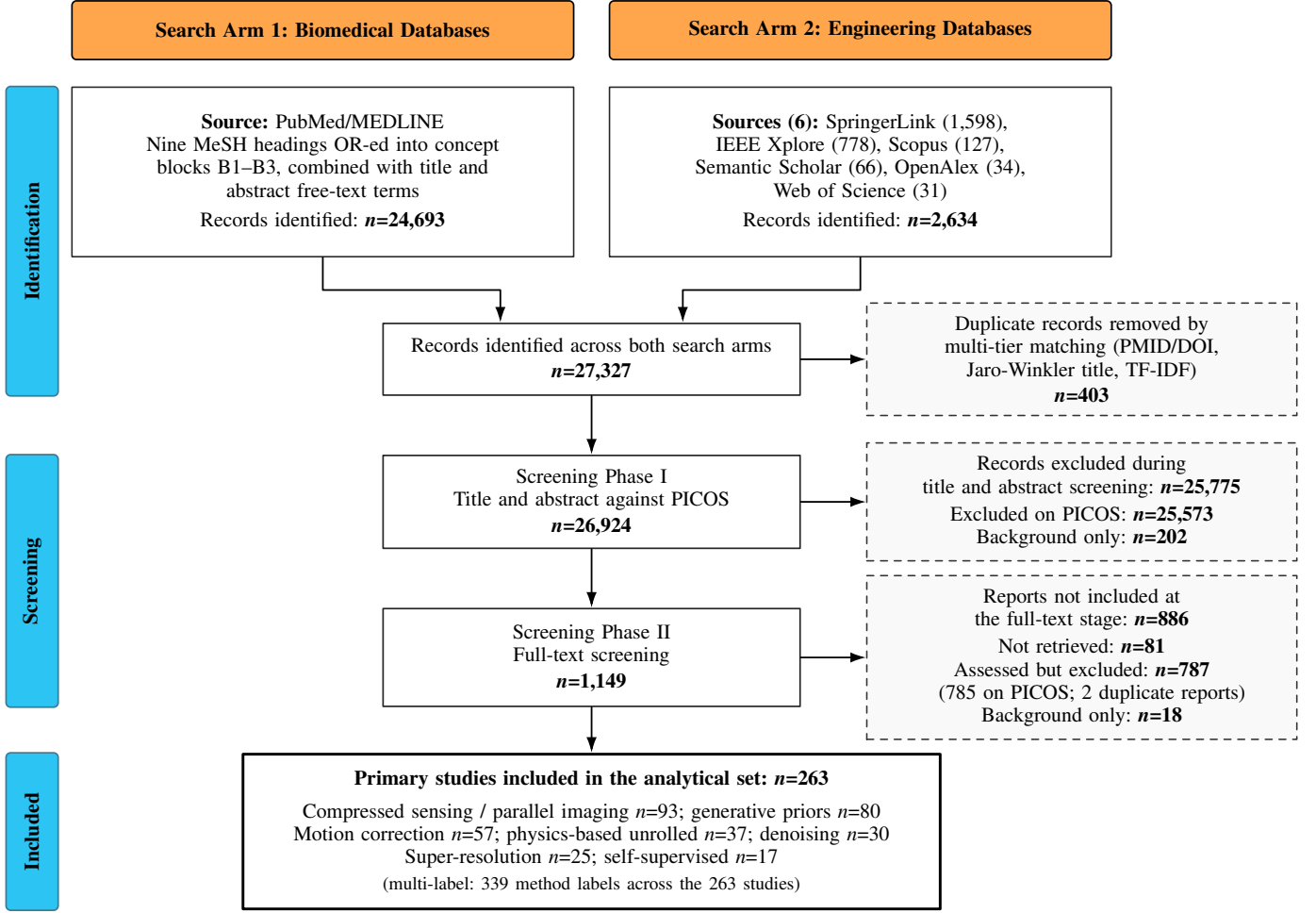
\begin{figure*}[!tbp]
\centering
\begingroup

\newcommand{\prcount}[1]{\mbox{\textit{n}=#1}}

\resizebox{\textwidth}{!}{%
\begin{tikzpicture}[
  line cap=round,
  line join=round,
  font=\small,
  header/.style={
    draw=black,
    line width=0.7pt,
    rounded corners=2pt,
    fill=orange!70,
    text width=7.3cm,
    minimum height=9mm,
    align=center,
    inner sep=2mm,
    font=\small\bfseries
  },
  sourcebox/.style={
    draw=black!85,
    line width=0.65pt,
    text width=7.3cm,
    minimum height=26mm,
    align=center,
    inner sep=2mm,
    fill=white
  },
  mainbox/.style={
    draw=black!85,
    line width=0.7pt,
    text width=6.0cm,
    minimum height=11mm,
    align=center,
    inner sep=2mm,
    fill=white
  },
  screeningbox/.style={
    draw=black!85,
    line width=0.7pt,
    text width=6.0cm,
    minimum height=15mm,
    align=center,
    inner sep=2mm,
    fill=white
  },
  exclusionbox/.style={
    draw=black!80,
    dashed,
    line width=0.65pt,
    text width=6.2cm,
    minimum height=12mm,
    align=center,
    inner sep=2mm,
    fill=gray!5
  },
  finalbox/.style={
    draw=black,
    line width=1.2pt,
    text width=10.2cm,
    minimum height=20mm,
    align=center,
    inner sep=2.5mm,
    fill=white
  },
  phasebar/.style={
    draw=cyan!50!black,
    line width=0.7pt,
    rounded corners=2pt,
    fill=cyan!65,
    minimum width=8mm,
    inner sep=0pt
  },
  flow/.style={
    -{Latex[length=2.4mm,width=1.6mm]},
    line width=0.75pt
  }
]

\node[header] (bioheader) at (4.95,0)
  {Search Arm 1: Biomedical Databases};

\node[header] (engheader) at (13.20,0)
  {Search Arm 2: Engineering Databases};

\node[sourcebox, below=4mm of bioheader] (bsource)
  {\textbf{Source:} PubMed/MEDLINE\\
   Nine MeSH headings OR-ed into concept\\
   blocks B1--B3, combined with title and\\
   abstract free-text terms\\[1mm]
   Records identified: \textbf{\prcount{24{,}693}}};

\node[sourcebox, below=4mm of engheader] (esource)
  {\textbf{Sources (6):} SpringerLink (1{,}598),\\
   IEEE Xplore (778), Scopus (127),\\
   Semantic Scholar (66), OpenAlex (34),\\
   Web of Science (31)\\[1mm]
   Records identified: \textbf{\prcount{2{,}634}}};

\coordinate (flowcenter) at
  ($(bioheader.center)!0.5!(engheader.center)$);

\coordinate (armbottom) at ([yshift=-10mm]bsource.south);

\node[mainbox, anchor=north] (identified) at (flowcenter |- armbottom)
  {Records identified across both search arms\\[0.5mm]
   \textbf{\prcount{27{,}327}}};

\node[exclusionbox, minimum height=16mm, right=10mm of identified] (duplicates)
  {Duplicate records removed by\\
   multi-tier matching (PMID/DOI,\\
   Jaro-Winkler title, TF-IDF)\\[0.5mm]
   \textbf{\prcount{403}}};

\node[mainbox, below=9mm of identified] (screened)
  {Screening Phase I\\
   Title and abstract against PICOS\\[0.5mm]
   \textbf{\prcount{26{,}924}}};

\node[exclusionbox, minimum height=18mm, right=10mm of screened] (abstractexcluded)
  {Records excluded during\\
   title and abstract screening:
   \textbf{\prcount{25{,}775}}\\[0.7mm]
   Excluded on \mbox{PICOS}: \textbf{\prcount{25{,}573}}\\
   Background only: \textbf{\prcount{202}}};

\node[screeningbox, below=9mm of screened] (fulltext)
  {Screening Phase II\\
   Full-text screening\\[0.5mm]
   \textbf{\prcount{1{,}149}}};

\node[
  exclusionbox,
  minimum height=18mm,
  right=10mm of fulltext
] (fulltextexcluded)
  {Reports not included at the full-text stage:
   \textbf{\prcount{886}}\\[0.7mm]
   Not retrieved: \textbf{\prcount{81}}\\
   Assessed but excluded: \textbf{\prcount{787}}\\
   \quad (785 on PICOS; 2 duplicate reports)\\
   Background only: \textbf{\prcount{18}}};

\node[finalbox, below=7mm of fulltext] (included)
  {\textbf{Primary studies included in the analytical set:
   \prcount{263}}\\[1.5mm]
   Compressed sensing / parallel imaging \prcount{93};
   generative priors \prcount{80}\\
   Motion correction \prcount{57};
   physics-based unrolled \prcount{37};
   denoising \prcount{30}\\
   Super-resolution \prcount{25};
   self-supervised \prcount{17}\\[0.5mm]
   \footnotesize (multi-label: 339 method labels across the 263 studies)};

\coordinate (joiny) at ([yshift=5mm]identified.north);

\draw[flow]
  (bsource.south) -- (bsource.south |- joiny)
  -| ([xshift=-14mm]identified.north);

\draw[flow]
  (esource.south) -- (esource.south |- joiny)
  -| ([xshift=14mm]identified.north);

\draw[flow] (identified) -- (screened);
\draw[flow] (screened) -- (fulltext);
\draw[flow] (fulltext) -- (included);

\draw[flow] (identified.east) -- (duplicates.west);
\draw[flow] (screened.east) -- (abstractexcluded.west);
\draw[flow] (fulltext.east) -- (fulltextexcluded.west);

\coordinate (phasex) at ([xshift=-6mm]bsource.west);

\coordinate (idtop) at (phasex |- bsource.north);
\coordinate (idbottom) at (phasex |- identified.south);
\node[phasebar,fit=(idtop)(idbottom)] (idbar) {};
\node[
  rotate=90,
  anchor=center,
  inner sep=0pt,
  font=\small\bfseries
] at (idbar.center) {Identification};

\coordinate (screentop) at (phasex |- screened.north);
\coordinate (screenbottom) at (phasex |- fulltext.south);
\node[phasebar,fit=(screentop)(screenbottom)] (screenbar) {};
\node[
  rotate=90,
  anchor=center,
  inner sep=0pt,
  font=\small\bfseries
] at (screenbar.center) {Screening};

\coordinate (inctop) at (phasex |- included.north);
\coordinate (incbottom) at (phasex |- included.south);
\node[phasebar,fit=(inctop)(incbottom)] (incbar) {};
\node[
  rotate=90,
  anchor=center,
  inner sep=0pt,
  font=\small\bfseries
] at (incbar.center) {Included};

\end{tikzpicture}%
}

\endgroup
\caption{PRISMA 2020 flow diagram for study selection across the two search
arms. Multi-tier deduplication reduced $27{,}327$ identified records to
$26{,}924$ unique records; title and abstract screening advanced $1{,}149$
candidates. Full texts were obtained for $1{,}068$ of them ($93.0\%$); $81$ could
not be retrieved. Following PRISMA 2020, exclusion and background retention are
recorded as distinct dispositions: of the $1{,}068$ reports assessed in full, $787$
were excluded ($785$ on explicit PICOS grounds and $2$ as duplicate reports of
included studies) and a further $18$ were retained as background material rather than
excluded, leaving $263$ primary studies in the analytical set
($263+787+18=1{,}068$).}
\label{fig:prisma_flow}
\end{figure*}

From 27{,}327 records identified across seven databases, multi-tier deduplication
removed 403 duplicates, leaving 26{,}924 unique records screened at title and abstract.
Of these, 1{,}149 proceeded to full-text eligibility, and full texts were obtained
for 1{,}068 of them ($93.0\%$; Section~\ref{sec:methods}). Each of
the 81 reports
still out of reach was checked against its publisher platform and assigned a recorded
reason: $53$ sit behind title-level subscriptions our institution does not hold, across
nineteen publishers; $13$ are Springer conference chapters sold only per chapter; $4$ are
ISMRM proceedings abstracts issued without a PDF; $4$ resolve to a whole proceedings volume
or to front matter rather than a single report; and $7$ carry no DOI and no locatable
source. Supplementary Table~S14 gives the per-publisher breakdown.
Access difficulty is itself informative, though it bounds the missing evidence rather than
characterising it. Of the $341$ assessed reports obtainable only through title-level
subscriptions and equivalent barriers, $294$ ($86.2\%$) proved ineligible on the same PICOS
criteria, $2$ were retained as background and $45$ were eligible ($294+2+45=341$; the $45$
resolve to $44$ studies, one being a preprint adjudicated a duplicate). Raw non-retrieval
counts therefore overstate the missing evidence by roughly sevenfold. But those $45$ were
concentrated in subscription clinical journals, and Section~\ref{res:gap_structure} shows
they differ systematically from the open literature in exactly the variable this review
measures, so $86.2\%$ is an upper bound on what non-retrieval costs us, not a demonstration
that it costs us nothing. Of the 1{,}068 assessed reports, 787 were
excluded (785 on PICOS grounds, 2 as duplicate reports of included studies) and 18
recorded as background, leaving \textbf{263 included
primary studies} spanning 1995--2026 (Fig.~\ref{fig:prisma_flow}). Each included study
was verified against the fullest text obtainable for it: complete full text for 258,
partial text for 1, and title, abstract and controlled vocabulary for 4
(Section~\ref{sec:methods}). Their aggregate
characteristics are given in Table~\ref{tab:characteristics}, and their thematic distribution across the research questions is summarized in Supplementary Table~S7; all 263 are listed, with full references, in Supplementary Tables~S1--S3.

\begin{table*}[!tbp]
\centering
\caption{Characteristics of the 263 included studies.}
\label{tab:characteristics}
\scriptsize
\setlength{\tabcolsep}{3.5pt}
\renewcommand{\arraystretch}{0.98}
\begin{tabular}{>{\raggedright\arraybackslash}p{0.20\textwidth} >{\raggedright\arraybackslash}p{0.24\textwidth} l >{\raggedright\arraybackslash}p{0.36\textwidth}}
\hline
\textbf{Demographic / Method Dimension} & \textbf{Category Breakdown} & \textbf{($n$)} & \textbf{Percentage / Key Methodological Notes} \\
\hline
\textbf{Publication Period} & 1995--2019 & $n=34$ & $12.9\%$ (Classical CS/PI \& early CNNs) \\
 & 2020--2022 & $n=55$ & $20.9\%$ (Unrolled physics \& GAN priors) \\
 & 2023--2026 & $n=174$ & $66.2\%$ (Diffusion priors \& 3D/4D foundation models)  \\
\hline
\textbf{Reconstruction Paradigm} \newline \emph{(multi-label; $339$ labels recorded)} & Compressed Sensing / Parallel Imaging & $n=93$ & Iterative sparsity \& coil sensitivity encoding \\
 & Generative (GAN/Diffusion/VAE) & $n=80$ & Adversarial perception \& posterior sampling \\
 & Motion Correction & $n=57$ & Intra-scan motion artifact suppression \\
 & Physics-based Unrolled & $n=37$ & Explicit $k$-space data consistency layers \\
 & Denoising & $n=30$ & Image-domain noise suppression \\
 & Super-resolution & $n=25$ & Resolution enhancement post-acquisition \\
 & Self-supervised / Unsupervised & $n=17$ & Training without fully sampled reference scans  \\
\hline
\textbf{MRI Pulse Sequences} \newline \emph{(multi-label; $490$ labels)} & T1-weighted (T1w) & $n=163$ & High-resolution anatomical cortical profiling \\
 & T2-weighted (T2w) & $n=122$ & Ventricular \& CSF pathology assessment \\
 & FLAIR & $n=85$ & Hyperintense peri-ventricular lesion detection \\
 & Diffusion-Weighted (DWI) & $n=72$ & Acute ischemic lesion (DWI core) evaluation \\
 & SWI / $T_2^*$-weighted & $n=48$ & Punctate cerebral microbleed hypointensity voids  \\
\hline
\textbf{Magnetic Field Strength} \newline \emph{(single-label; predominant field where several are used)} & 3.0 Tesla ($3\text{T}$) & $n=153$ & $58.2\%$ (Primary clinical scanner benchmark) \\
 & 7.0 Tesla ($7\text{T}$) & $n=35$ & $13.3\%$ (Ultra-high-field research acquisitions) \\
 & 1.5 Tesla ($1.5\text{T}$) & $n=21$ & $8.0\%$ (Routine clinical throughput imaging) \\
 & \emph{Not stated} & $n=50$ & $19.0\%$ (Full text obtained; no field strength stated) \\
 & \emph{Not ascertainable} & $n=4$ & $1.5\%$ (Only abstract or partial text available)  \\
\hline
\textbf{Evaluation Paradigm} \newline \emph{(multi-label; $177$ labels recorded across $144$ studies)} & Pixel-wise (PSNR/\allowbreak SSIM/\allowbreak NRMSE) & $n=84$ & $47.5\%$ of recorded (legacy fidelity reporting) \\
 & Observer / Reader Studies & $n=57$ & $32.2\%$ of recorded (task-based reader assessment; verified subset) \\
 & Uncertainty Estimation & $n=18$ & $10.2\%$ of recorded (posterior variance maps) \\
 & Learned / Perceptual & $n=18$ & $10.2\%$ of recorded (rank-trained metrics)  \\
\hline
\multicolumn{4}{>{\raggedright\arraybackslash}p{0.92\textwidth}}{\scriptsize\textbf{Co-reporting.} Of the $144$ studies with a recorded evaluation paradigm, $28$ report more than one. Critically, only $\mathbf{18}$ studies in the entire corpus ($6.8\%$ of $263$; $12.5\%$ of the $144$ recorded) report \emph{both} a pixel-wise fidelity metric \emph{and} a reader assessment on the same data, which is the minimum precondition for computing metric-versus-reader agreement on which safety claims implicitly depend. See Section~\ref{sec:results}.} \\
\hline
\end{tabular}
\end{table*}
 Corpus growth is recent: 34 studies
appeared in 1995--2019, 55 in 2020--2022 and 174 in 2023--2026.

\subsection{RQ1: the metric--reader comparison is almost never made, and where readers appear they agree only partially}
Across the 84 studies reporting objective fidelity metrics and the 57 verified reader studies,
the comparisons that actually exist show agreement between fidelity metrics and
radiologist assessment that is partial and condition-dependent
(Table~\ref{tab:rq1_synthesis}). The size of that evidence base should be stated in the
text and not left to the table caption: only two of the four rows in
Table~\ref{tab:rq1_synthesis} were collected on reconstructions of undersampled
$k$-space. \citet{tang2025incorporating} collected radiologist rankings on synthetically
corrupted image pairs, and \citet{sommer2020correction} on clinical scans corrupted by
real patient motion rather than by undersampling. The table that carries RQ1 therefore
rests on two on-target rows, which is the honest size of the direct evidence for the
metric--reader comparison in this setting. In the fastMRI 2020 challenge's reader phase, six
radiologists ranked submissions for quality of pathology depiction; their concordance,
by Kendall's coefficient, was $0.457$ in the $4\times$ track and $0.386$ in the
$8\times$ track, rising to $0.781$ only in the transfer track, and the organisers
report that concordance improved as SSIM scores diverged, which is to say the metric
separated methods reliably only where their outputs already differed grossly
\citep{muckley2021results}. When radiologist preference is instead used as the training
signal, purpose-built quality networks predict expert rankings at $75.2\%\pm1.3\%$ and
$79.2\%\pm1.7\%$, against radiologists' own interobserver agreement of
$61.5\%\pm5.5\%$; the same study finds rank-based reading far more reliable than Likert
scoring ($70.4\%$ versus $25\%$ agreement between radiologists)
\citep{tang2025incorporating}. Reader tolerance of acceleration is real but bounded: in
a retrospective evaluation spanning accelerations up to $100\times$, $94\%$ of volumes
($65$ of $69$) were rated sufficient for potential screening use at up to $14\times$
(interreader intraclass correlation $0.875$), with the threshold for full diagnostic
quality lower still \citep{radmanesh2022exploring}. The one direct clinical observation of the failure this review is about comes from an
emergency-department and inpatient comparison, and it is a single case. Across $66$
paired examinations in which a two-minute ultrafast protocol (T1w, T2/T2$^{*}$, FLAIR
and DWI, machine-learning-assisted reconstruction) was read against a ten-minute
reference protocol, two blinded neuroradiologists agreed on the main clinical diagnosis
in $65$ of $66$ cases ($98.5\%$), rated overall diagnostic quality no different between
protocols ($P>.05$), preferred the reference protocol for image noise and geometric
distortion ($P<.05$) and preferred the accelerated protocol for reduced motion artifact
($P<.01$) \citep{lang2024ultrafast}. The single discrepant case is the one that matters
here: a punctate focus of restricted diffusion, read as possible acute-to-subacute
infarct on the reference DWI, was less conspicuous on the accelerated DWI, which the
authors attribute to a combination of image-quality difference and section positioning,
the lesion being smaller than the section thickness ($4\text{ mm}$ reference against
$5\text{ mm}$ accelerated) and so subject to partial-volume contamination. They draw the
inference themselves, that the protocol's sensitivity to small findings is an open
question. One case in $66$ is not a prevalence estimate and we do not offer it as one.
It is a worked instance of the pattern this review predicts: an aggregate quality
comparison that finds no difference, and inside it a focal finding that a reader saw
and the aggregate did not. Where readers assess artifacts
directly, blinded reading detects what reference-based metrics cannot express: a
two-neuroradiologist blinded study of 28 clinical scans with real patient motion found
significant artifact-score reductions after network correction ($p<.03$), a setting in
which the motion-corrupted reference invalidates PSNR and SSIM by construction
\citep{sommer2020correction}. Motion also interacts with the sampling scheme itself,
appearing as ghosting under Cartesian trajectories and as streaking or blurring under
radial ones, so the visual form of the same physical corruption differs by
protocol \citep{schauman2026an}. The reference pipeline is not neutral either:
conventional SWI homodyne phase filtering has been shown to create punctate artifactual
microhaemorrhage-like foci, so processing can manufacture findings as well as remove
them \citep{li2015artifactual}.

No included study reports a rank correlation between a pixel-wise fidelity metric and
radiologist diagnostic assessment for deep learning-based brain reconstruction, and none
evaluated a model observer. The second statement needs a boundary that the first does
not. Our search vocabulary (Table~\ref{tab:queries}, block B3) contains reader-study and
metric terms but no model-observer vocabulary, so a study that used a channelised
Hotelling or non-prewhitening observer and described it only in those terms could have
been missed at retrieval. What we can say is that none of the $263$ studies retrieved,
screened and extracted here reports one, and that the term does not appear in any
extracted evaluation label; whether the true field-wide count is zero or merely very
small, it is not a paradigm this corpus can be said to use. That absence is itself the RQ1 finding: only $18$ of
$263$ studies ($6.8\%$, a screening-level lower bound; Section~\ref{sec:methods})
record a fidelity metric and a reader assessment on the same data, so the field's
central proxy relationship, that high fidelity scores imply diagnostic adequacy, is
largely unmeasured rather than merely disputed.

\begin{table*}[!tbp]
\centering
\caption{What the reader studies in this corpus actually measured about metric--reader
agreement (RQ1). Every value is quoted from the named study, with the reported
coefficient and its definition stated. No included study reports a metric--reader rank
correlation for deep learning-based brain reconstruction; the rows below are the closest
measurements that exist, and the second column states for each row what the readers were
actually shown, because two of the four rows were not collected on reconstructions of
undersampled $k$-space.}
\label{tab:rq1_synthesis}
\small
\setlength{\tabcolsep}{5pt}
\renewcommand{\arraystretch}{1.18}
\begin{tabular}{>{\raggedright\arraybackslash}p{0.16\textwidth} >{\raggedright\arraybackslash}p{0.24\textwidth} >{\raggedright\arraybackslash}p{0.20\textwidth} >{\raggedright\arraybackslash}p{0.26\textwidth}}
\hline
\textbf{Study} & \textbf{Readers and design} & \textbf{Comparison performed} & \textbf{Reported quantity (as defined in the study)} \\
\hline
fastMRI 2020 challenge \citep{muckley2021results} & 6 radiologists ranking submissions for quality of pathology depiction & Radiologist concordance across accelerations; SSIM-based ranking vs reader ranking & Kendall's $W$ $= 0.457$ ($4\times$), $0.386$ ($8\times$), $0.781$ (transfer track); concordance improved as SSIM scores diverged \\
\hline
\citet{tang2025incorporating} & 7 radiologists, rank-based reading; 2 also Likert scoring. \textbf{Rankings were collected on synthetically corrupted image pairs, not on reconstructions of undersampled $k$-space} & Quality networks trained on radiologist rankings vs MSE/SSIM as optimisation targets & Ranking accuracy $75.2\%\pm1.3\%$ and $79.2\%\pm1.7\%$ with reference image; radiologist interobserver agreement $61.5\%\pm5.5\%$; ranking vs Likert agreement $70.4\%$ vs $25\%$ \\
\hline
\citet{radmanesh2022exploring} & Retrospective; neuroradiologist scoring of diagnostic quality at accelerations up to $100\times$ & Reader-assigned acceleration thresholds for diagnostic and screening quality & $65/69$ volumes ($94\%$) sufficient for potential screening at $\le 14\times$; interreader ICC $0.875$; mean absolute interreader difference in threshold acceleration factor $0.7\times$ (level 1) and $5.4\times$ (level 2) \\
\hline
\citet{sommer2020correction} & 2 neuroradiologists, blinded, 28 clinical cases with real motion & Artifact severity (5-point scale) before vs after network correction & Significant reduction in mean artifact scores ($p<.03$); reference-based metrics inapplicable, as the reference itself is motion-corrupted \\
\hline
\end{tabular}
\end{table*}

Table~\ref{tab:acceleration_spectrum} reports how evaluation practice varies with the
acceleration factor actually stated in each study. It is built only from what the
corpus records: the acceleration factor was extracted post hoc from the full texts of
all 263 included studies with a verbatim supporting quotation for every value, and
$63$ studies ($24.0\%$) state one. Studies are grouped by the highest factor at which
they evaluate their method. The table deliberately carries no fidelity threshold,
reader-score difference or metric--reader rank statistic per acceleration band,
because no included study reports one: that absence is the finding of
Section~\ref{sec:results}, not an omission from the table.

\begin{table*}[!tbp]
\centering
\caption{Evaluation practice by the highest acceleration factor ($R$) each study
reports, among the $63$ of $263$ included studies that state one. Bands are half-open,
so a study reporting exactly $R=4$ falls in the second band and one reporting exactly
$R=8$ in the third. Counts are studies; percentages are within band. \emph{Reader}, any observer or reader assessment recorded;
\emph{M$+$R}, a pixel-wise fidelity metric and a reader assessment recorded on the same
data; \emph{Halluc.}, hallucination or fidelity instability among the recorded failure
modes; \emph{Artifacts}, aliasing, ghosting or Gibbs artifacts among them. Acceleration
factors were extracted from full text after the registered extraction round, each with a
verbatim supporting quotation and excluding matches inside reference lists and citation
contexts; the $407$ retained quotations come from the $63$ studies that state a factor,
a mean of $6.5$ quotations per such study, and are released as
\texttt{acceleration\_factors\_evidence.csv}. The extraction is screening-level, so
counts are lower bounds. No study at any acceleration factor reports a rank correlation
between a fidelity metric and reader judgement.}
\label{tab:acceleration_spectrum}
\footnotesize
\setlength{\tabcolsep}{2.5pt}
\renewcommand{\arraystretch}{1.2}
\begin{tabular}{l c c c c c >{\raggedright\arraybackslash}p{0.219\textwidth}}
\hline
\textbf{Acc. ($R$)} & \textbf{Studies} & \textbf{Reader} & \textbf{M$+$R} &
\textbf{Halluc.} & \textbf{Artifacts} & \textbf{Sequences most often evaluated} \\
\hline
$2\leq R<4$      & $32$ & $7$ ($21.9\%$) & $1$ ($3.1\%$)  & $7$ ($21.9\%$) & $25$ ($78.1\%$) & T1w 15, T2w 12, FLAIR 11 \\
$4\leq R<8$      & $17$ & $5$ ($29.4\%$) & $1$ ($5.9\%$)  & $2$ ($11.8\%$) & $12$ ($70.6\%$) & T1w 9, T2w 9, FLAIR 7 \\
$R\geq 8$        & $14$ & $5$ ($35.7\%$) & $4$ ($28.6\%$) & $2$ ($14.3\%$) & $8$ ($57.1\%$)  & T1w 9, T2w 8, FLAIR 8 \\
\hline
\textbf{All} & $\mathbf{63}$ & $\mathbf{17}$ ($\mathbf{27.0\%}$) & $\mathbf{6}$ ($\mathbf{9.5\%}$) & $\mathbf{11}$ ($\mathbf{17.5\%}$) & $\mathbf{45}$ ($\mathbf{71.4\%}$) & T1w 33, T2w 29, FLAIR 26 \\
\hline
\multicolumn{7}{>{\raggedright\arraybackslash}p{0.86\textwidth}}{\footnotesize The $R$-reporting subset is close to
the whole corpus on reader assessment ($27.0\%$ against $21.7\%$ for all $263$), so it is
not a systematically better-evaluated stratum. Reader assessment rises with acceleration
($21.9\%\rightarrow29.4\%\rightarrow35.7\%$) and paired metric--reader reporting
concentrates in the most accelerated band ($4/14$ against $1/32$ in the $2\leq R<4$ band), but
the counts are small and most studies in every band (including $9$ of the $14$ with $R\geq 8$) report no reader assessment at all. Artifact description is the most
frequently recorded failure mode at every acceleration level, whereas hallucination or
fidelity instability is recorded in $11$ of $63$.} \\
\hline
\end{tabular}
\end{table*}

\subsection{RQ2: four mechanisms explain the divergence}
Four mechanisms account for why the metrics miss what the readers see.
Table~\ref{tab:mechanisms} states each one formally alongside the clinical failure it
permits, and Supplementary Table~S8 gives the fuller derivations and evidence sources.

\begin{table*}[!tbp]
\centering
\caption{The four mechanisms by which a whole-volume fidelity score fails to register a
focal diagnostic error (RQ2). The third column names a failure the mechanism permits,
not one this review has observed at a stated frequency; the fourth states why the score
does not move. Numerical entries are computed from
Equation~\eqref{eq:dpsnr_identity} under the stated error model, not measured in any
primary study.}
\label{tab:mechanisms}
\footnotesize
\setlength{\tabcolsep}{2pt}
\renewcommand{\arraystretch}{1.18}
\begin{tabular}{>{\raggedright\arraybackslash}p{0.16\textwidth} >{\raggedright\arraybackslash}p{0.26\textwidth} >{\raggedright\arraybackslash}p{0.255\textwidth} >{\raggedright\arraybackslash}p{0.275\textwidth}}
\hline
\textbf{Mechanism} & \textbf{Formal statement} & \textbf{Clinical failure it permits} & \textbf{Why the score does not move} \\
\hline
(a) Spatial error averaging &
$L_2$ error is averaged uniformly over all $H\times W\times D$ voxels, so a lesion
contributes with weight $f = V_{\text{lesion}}/V_{\text{total}}$ &
Erasure of an acute lacunar infarct on DWI or FLAIR, or of a punctate microhaemorrhage
on SWI or $T_2^{*}$ &
By \eqref{eq:dpsnr_identity}, erasing a $100\text{ mm}^3$ infarct at $r=100$ moves global
PSNR by $0.027$--$0.036\text{ dB}$; a $5\text{ mm}^3$ microhaemorrhage by
$1.3$--$1.8\times10^{-3}\text{ dB}$ \\
\hline
(b) Reference dependence &
Every reference-based score presupposes a registered, artifact-free reference image &
A valid motion-corrected reconstruction is penalised against a corrupted reference, and
genuine correction is scored as error &
The comparison is no longer between reconstruction and truth; PSNR and SSIM become
uninterpretable rather than merely noisy \\
\hline
(c) Loss symmetry &
$(\hat{X}_i - X_i)^2 = (X_i - \hat{X}_i)^2$: the objective is indifferent to the sign of
the error &
A suppressed hyperintense lesion is scored identically to added background noise of the
same magnitude &
Diagnosis is asymmetric where the metric is not: a missed finding and a spurious
brightening carry the same penalty \\
\hline
(d) Realism--reward trade-off &
Adversarial and perceptual objectives reward outputs typical of the training
distribution, $\log D_\phi(\hat{X})$, not outputs faithful to the measurement &
Synthesis of plausible vessels, sulci or texture that was never acquired &
Perceptual scores can rise as measurement fidelity falls, so the objective pays for the
hallucination it should penalise \\
\hline
\end{tabular}
\end{table*}

\emph{Spatial averaging.} Conventional $L_2$ metrics average residual error uniformly
over all $H\times W\times D$ voxels, so the total decomposes over lesion and background
regions:
\predisplaypenalty=0
\begin{equation}
\begin{aligned}
\text{MSE}(\hat{X}, X) &= \frac{V_{\text{lesion}}}{V_{\text{total}}}
\text{MSE}_{\text{lesion}} \\
&\quad + \left(1 - \frac{V_{\text{lesion}}}{V_{\text{total}}}\right)
\text{MSE}_{\text{bg}}.
\end{aligned}
\end{equation}
\predisplaypenalty=10000
With lesion volume fraction $f = V_{\text{lesion}}/V_{\text{total}}$ and local error ratio
$r = \text{MSE}_{\text{lesion}}/\text{MSE}_{\text{bg}}$, the global PSNR change caused by a
localized edit follows directly:
\begin{equation}
\label{eq:dpsnr_identity}
|\Delta\mathrm{PSNR}| = 10\log_{10}\left(1 + f(r - 1)\right) \quad (\text{dB}).
\end{equation}
The consequence is quantitative rather than rhetorical, and it is worth anchoring on a
lesion a neuroradiologist would expect to resolve rather than on the smallest
arithmetically convenient one. An acute lacunar infarct of $100\text{ mm}^3$ is a sphere
$5.8\text{ mm}$ across: on a clinical DWI acquired at $4$--$5\text{ mm}$ section
thickness \citep{lang2024ultrafast} that is several voxels wide in-plane and spans one
to two sections, so it is resolved, if only just, in the through-plane direction. It
occupies a volume fraction
$f = 6.3$--$8.3\times10^{-5}$ of a $1.2$--$1.6\text{ L}$ brain. Even at a hundredfold
local error ratio ($r=100$), Equation~\eqref{eq:dpsnr_identity} gives
$|\Delta\mathrm{PSNR}| = 0.027$--$0.036\text{ dB}$. Erasing a resolvable acute infarct
outright therefore moves global PSNR by about three hundredths of a decibel, which is
the same order as the third decimal place of a score conventionally reported to two.

\begin{figure*}[!tbp]
\centering
\includegraphics[width=0.97\textwidth]{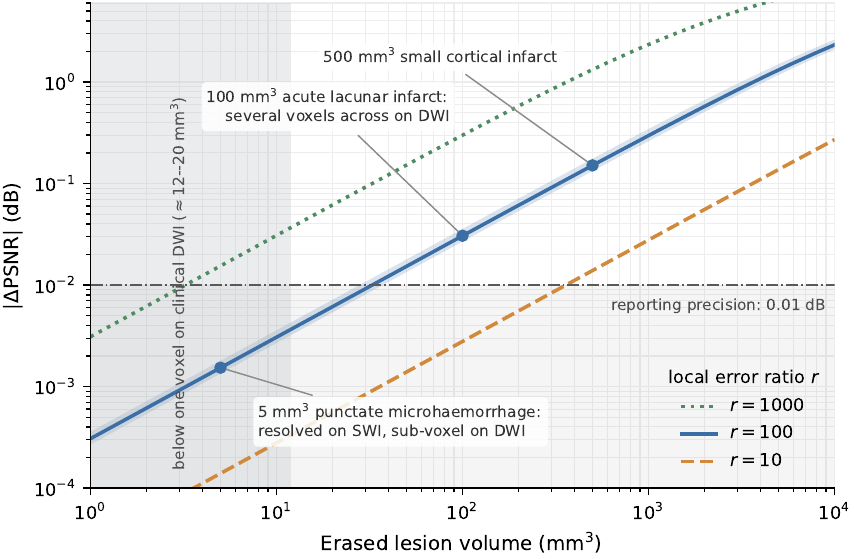}
\caption{The PSNR cost of erasing a focal lesion, from
Equation~\eqref{eq:dpsnr_identity}. Curves give $|\Delta\mathrm{PSNR}|$ against erased
lesion volume for three local error ratios $r$ at a $1.4\text{ L}$ brain; the shaded band
around $r=100$ spans the $1.2$--$1.6\text{ L}$ range. Markers show the three clinical
anchors used in the text. The horizontal line at $0.01\text{ dB}$ is the two-decimal
precision at which PSNR is conventionally reported, and the shaded left-hand region is
the volume below a single voxel on a clinical DWI acquisition, where erasure is not
defined because the finding is not resolved. Every value is computed from the identity
under the stated error model; nothing in this figure is measured.}
\label{fig:dpsnr_identity}
\end{figure*}

The limiting case is worth stating precisely, because it is easy to state loosely. A
punctate cerebral microhaemorrhage of $5\text{ mm}^3$ is $2.1\text{ mm}$ across. It is
\emph{not} resolved on clinical DWI, whose voxels are roughly $12$--$20\text{ mm}^3$, so
no reconstruction can be said to erase it there; it is not resolved in the first place.
It is several voxels across on the high-resolution SWI or $T_2^{*}$-weighted acquisition
on which microhaemorrhages are actually read, and on that sequence its erasure costs
$f = 3.1$--$4.2\times10^{-6}$ and $|\Delta\mathrm{PSNR}| = 1.3$--$1.8\times10^{-3}\text{
dB}$, below the precision at which the score is reported at all. Scaling in the other
direction, a $500\text{ mm}^3$ cortical infarct costs $\approx 0.15\text{ dB}$
($0.13$--$0.18\text{ dB}$ across the same brain-volume range).
Fig.~\ref{fig:dpsnr_identity} plots the identity across this range, with the
sub-voxel region marked. The argument does not need the extreme case: it holds by one
to two orders of magnitude at a lesion size that is unambiguously resolved and
unambiguously matters. These are computed bounds from the stated $f$ and $r$, not
measurements from any primary study, and the error model behind them, a uniform
background error and a free local error ratio, is an assumption we state rather than a
property of any reconstruction.

\emph{Loss-function symmetry.} MSE and SSIM are indifferent to the sign of the error,
since $(\hat{X}_i - X_i)^2 = (X_i - \hat{X}_i)^2$, whereas diagnosis is not: suppressing
a true hyperintense lesion means the finding is missed, while adding background noise
costs nothing clinically. An evaluation respecting that asymmetry must penalise the two
directions differently, for example
\begin{equation}
\begin{aligned}
\mathcal{L}_{\text{asym}}(\hat{X}, X) &= \alpha \sum_{i \in \Omega_{\text{lesion}}}
\max(0, X_i - \hat{X}_i)^2 \\
&\quad + \beta \sum_{i \in \Omega_{\text{bg}}} (\hat{X}_i - X_i)^2,
\quad (\alpha \gg \beta).
\end{aligned}
\end{equation}
Several included studies already build objectives of this shape: lesion-mask-weighted
losses motivated explicitly by the dilution of lesion gradients among background voxels
\citep{junhyeok2026lesionaware}, overlap-weighted reconstruction losses that up-weight
structure against background
\citep{nivetha2023sadir}, and gradient-based structure-preservation terms
\citep{zhang2026iterative}. Their existence is evidence that the symmetry problem is
recognised in practice, though none of these studies pairs the correction with a reader
assessment.

\emph{Realism--reward trade-off.} Adversarial and perceptual objectives reward outputs
that look like the training distribution, which is not the same as outputs faithful to
the measurement, so a generative reconstruction can gain plausible texture that was
never acquired \citep{zhou2024adversarial,bhadra2021hallucinations}.

\emph{Reference dependence.} All four metrics presuppose a registered, artifact-free
reference. Intra-scan motion invalidates that assumption, and $69$ of $263$ studies
rest their reference standard solely on a fully sampled acquisition, which is valid only
where the subject did not move.

\subsection{RQ3: what current evaluation paradigms can and cannot establish}
\label{res:rq3}
Read against those mechanisms, the corpus supports a paradigm-by-paradigm account of
what each evaluation approach can demonstrate about prior-induced diagnostic error.
Table~\ref{tab:rq3_metrics} states the account compactly; the argument behind it is
given here, because two of its six rows are analytical claims we derive rather than
findings we extracted, and an analysis presented only in a table has not been argued.

\emph{What the four empirical paradigms establish.} Pixel-wise fidelity metrics
establish aggregate agreement with a reference and nothing about focal change: by
\eqref{eq:dpsnr_identity} the penalty for erasing a resolvable acute infarct is of the
order of the reporting precision, so a method can pass any threshold anyone would set
while failing on the finding that matters. Perceptual and rank-trained metrics track
reader preference better than MSE or SSIM as an optimisation target, but they inherit
the variability of the readers they were trained on and, in the one study of this kind
in the corpus, were calibrated on synthetically corrupted images rather than on
reconstructions of undersampled $k$-space \citep{tang2025incorporating}, so their
transfer to the reconstruction setting is assumed rather than shown. Uncertainty
estimates flag where a model is unsure, which is not the same question: a prior
confident enough to remove a lesion is confident in the region where it removed it, so
posterior variance can be low exactly where the diagnostic error is. Task-based
observer assessment does settle the question directly, and it is the only paradigm here
that does, which is why its scarcity in this corpus (57 of 263 studies, and 18 paired
with a metric) is the review's central finding rather than an incidental one.

\emph{Two limits that follow from the physics, not from the corpus.} The first concerns
the check that reconstruction papers most often use to argue that an output is faithful
to the data: the measurement-space residual
$\|\mathbf{A}\hat{X} - Y\|_2/\|Y\|_2$ against a tolerance. This check cannot detect
prior-induced edits at all, and the reason is structural rather than statistical. Write
the reconstruction as the sum of its measured and unmeasured components,
$\hat{X} = \mathbf{A}^{+}\mathbf{A}\hat{X} + (\mathbf{I}-\mathbf{A}^{+}\mathbf{A})\hat{X}$.
The second term lies in the null space $\mathcal{N}(\mathbf{A})$, and $\mathbf{A}$
annihilates it by definition, so any amount of structure written there, an erased lesion
or an invented vessel alike, leaves the residual exactly unchanged. Undersampling is
what creates that null space; the more aggressive the acceleration, the larger the
subspace in which a learned prior can write freely without moving the number the field
uses to certify data consistency. A residual check is therefore a two-sided
measurement-fidelity band and not a safety test, and its apparent reassurance grows
least trustworthy precisely where acceleration is highest.

The second limit is the constructive corollary. If prior-supplied content lives in
$\mathcal{N}(\mathbf{A})$, then isolating $(\mathbf{I}-\mathbf{A}^{+}\mathbf{A})\hat{X}$
and inspecting it is the natural place to look for erasure and hallucination, since that
is where both must reside \citep{bhadra2021hallucinations,gottschling2020troublesome}. This is
the one paradigm in Table~\ref{tab:rq3_metrics} that is not represented in the corpus at
all, and its practical obstacle is the same one that limits third-party auditing
generally: computing the projection requires the forward operator, hence the raw
$k$-space and the coil sensitivities, which commercial inline pipelines do not export
(Section~\ref{sec:discussion}). Causal and counterfactual auditing, which asks the most
direct question of all, whether a specific finding survives a controlled change of
acquisition, appears in a small part of the corpus and almost never for
physics-constrained $k$-space reconstruction
\citep{pawlowski2020deep,gngr2023adaptive,wei2026latent}.

The answer to RQ3 is therefore not that current evaluation is weak but that it is
mismatched: each paradigm in use answers a question adjacent to the safety question,
and no paradigm now in use establishes reconstruction safety on its own. The two that
could, task-based observer assessment and null-space inspection, are respectively rare
and absent, and the requirements in Section~\ref{sec:requirements} follow from that
pairing.

\begin{table*}[!tbp]
\centering
\caption{What each evaluation paradigm can and cannot establish about prior-induced diagnostic error (RQ3). Rows 1--4 summarize evaluation practice and performance as reported in the included studies; quoted values are representative results from the cited primary studies, not pooled estimates. Rows 5--6 state analytical limits derived by the authors from the acquisition physics (our own analysis; Section~\ref{sec:results}) and are labelled as such rather than attributed to any primary study.}
\label{tab:rq3_metrics}
\scriptsize
\setlength{\tabcolsep}{3pt}
\renewcommand{\arraystretch}{1.0}
\begin{tabular}{>{\raggedright\arraybackslash}p{0.1617\textwidth} >{\raggedright\arraybackslash}p{0.1813\textwidth} >{\raggedright\arraybackslash}p{0.4318\textwidth} >{\raggedright\arraybackslash}p{0.147\textwidth}}
\hline
\textbf{Evaluation Paradigm} & \textbf{Core Mechanism} & \textbf{What It Can and Cannot Establish} & \textbf{Key References} \\
\hline
1. Pixel-wise fidelity metrics & $L_2$ / SSIM whole-image voxel averaging & Aggregate agreement with a reference only. Cannot resolve focal change: by \eqref{eq:dpsnr_identity}, erasing a resolvable $100\text{ mm}^3$ lacunar infarct at $r=100$ costs $0.03\text{ dB}$. Requires a registered reference & Recorded in $84$ of the $144$ evaluation-labelled studies \\
\hline
2. Perceptual \& learned metrics & Networks trained on radiologist rankings & Tracks reader preference better than MSE/SSIM as an optimisation target ($75.2\%\pm1.3\%$ and $79.2\%\pm1.7\%$ against readers' own $61.5\%\pm5.5\%$). Inherits training-reader variability, and was calibrated on synthetically corrupted data rather than clinically undersampled acquisitions, so transfer is assumed (Section~\ref{res:rq3}) & \citet{tang2025incorporating} ($n=18$ in corpus) \\
\hline
3. Uncertainty estimation & Posterior sampling / Bayesian variance $p(X \mid Y)$ & Flags out-of-distribution inputs and regions of low confidence. Does not attribute missing structure to prior vs sampling operator; high confidence in hallucinated regions is possible & \citet{khawaled2024npbrec} ($n=18$ in corpus) \\
\hline
4. Task-based observer assessment & Human reader performance on a diagnostic task; model observers are its automated form, unrepresented here \citep{barrett1993model,barrett2004foundations,icru54} & Establishes diagnostic sufficiency directly; the gold standard in assessment theory. Costly and reader-variable ($61.5\%\pm5.5\%$, falling to $25\%$ under absolute Likert scoring), and rarely paired with a metric: only $18/263$ report both & \citet{radmanesh2022exploring}, \citet{lang2024ultrafast} ($n=57$ in corpus) \\
\hline
5. Measurement-space residual check & $\|\mathbf{A}\hat{X} - Y\|_2 / \|Y\|_2$ against tolerance & \textbf{Cannot detect prior-induced edits}: the residual is invariant to anything in $\mathcal{N}(\mathbf{A})$. A two-sided measurement-fidelity band, not a safety test & Common diagnostic check in the reviewed studies \\
\hline
6. Null-space / hallucination mapping & Structure written into the null-space component $(\mathbf{I}-\mathbf{A}^{+}\mathbf{A})\hat{X}$ & Isolates prior-supplied content, where erasure and hallucination must reside. Requires $\mathbf{A}$, hence raw $k$-space and coil sensitivities & \citet{bhadra2021hallucinations}, \citet{gottschling2020troublesome}; formalization for reconstruction auditing is author-derived \\
\hline
\end{tabular}
\end{table*}

\subsection{Structure of the evaluation gap}
\label{res:gap_structure}
The counts above establish that safety-relevant evaluation is scarce. Analysis of the
same per-study records establishes where the scarcity comes from, and indicates a
division of labour rather than uniform neglect.

\emph{The benchmarks cannot show the failure the field is worried about.} One hundred
and nine of
the 263 studies ($41.4\%$) name a public dataset, and across the whole corpus those
references resolve to eleven datasets: fastMRI, the Human Connectome Project, ADNI, IXI,
UK Biobank, BraTS, OASIS, Calgary--Campinas, M4Raw, the developing Human Connectome
Project and TCGA. Dataset labels are multi-valued
(136 cohort assignments across the 109 studies), and of the 109, the cohorts named are
healthy volunteers or population samples (55, $50.5\%$), chronic neurodegenerative
cohorts (31, $28.4\%$), brain tumour cohorts (16, $14.7\%$), or raw-$k$-space
benchmarks not curated for pathology (34, $31.2\%$). These four shares are multi-label
and sum to $124.8\%$, a study naming two cohorts being counted in both; every other
percentage here is single-label. \emph{Within this corpus, no study
benchmarks against a dataset curated for acute ischemic stroke or intracranial
haemorrhage.} The sharper version of this finding sits in the annotation layer the
field has already built. fastMRI+ supplies $7{,}570$ expert bounding-box annotations
across thirty brain pathology categories \citep{zhao2022fastmriplus}, and not one of the
thirty is a haemorrhage category of any kind, microhaemorrhage included. The apparent
counterexample should be stated plainly rather than left for a reader to find, because it
is the category closest to this article's own worked example: fastMRI+ \emph{does} carry
an image-level \emph{Lacunar Infarct} label ($113$ annotations across $32$ subjects),
alongside study-level \emph{Small Vessel Chronic White Matter Ischemic Change} and a
\emph{Global Ischemia} label recorded for a single subject. It does not rescue the
benchmark, and the reasons are the ones that matter for reconstruction safety. No
fastMRI+ category distinguishes acute from chronic infarction, so a lacunar label carries
no acuity and cannot identify the acute presentation whose erasure is at issue. The
limit is in the source images as much as in the label set: the annotated series are
axial T2-FLAIR, T1w or contrast-enhanced T1w only, so neither DWI, on which infarct
acuity is established, nor SWI or $T_2^{*}$-weighted imaging, on which microhaemorrhage
is read, was annotated at all. Acute vascular pathology is therefore absent even where
annotation infrastructure exists, and it is absent for a reason that adding labels alone
would not repair. Supplementary Table~S10 sets out the full derivation, including the
sub-selection of $1{,}001$ of the $5{,}847$ fastMRI brain studies, the category listing,
and the single-annotator limitation that travels with these annotations. Stroke-lesion
datasets do exist, ISLES
\citep{petzsche2022isles} and ATLAS \citep{liew2022atlas} among them, but they are
segmentation benchmarks distributed as processed images without raw $k$-space, so they
cannot serve reconstruction evaluation as released. This matters mechanically: lesion
erasure cannot be measured on cohorts that contain no acute lesions, so a model can
lead every public leaderboard in the field while remaining untested against the
failure mode that would matter most in an emergency department. Where a
pathology-labelled dataset is used at all it is BraTS, whose tumours are large,
high-contrast mass lesions, close to the easiest possible case for lesion preservation
and the least informative about the lacunar and punctate findings at issue here.
The benchmark infrastructure and the metric therefore fail in the same direction.

\emph{Rigour is divided between two literatures.} Of the 263 studies, 84 ($31.9\%$)
release code but report no reader assessment, 44 ($16.7\%$) report a reader assessment
but release no code, 122 ($46.4\%$) do neither, and only 13 ($4.9\%$) do both. The
division tracks publication venue. Applying the post-hoc classification rule stated
in Section~\ref{sec:methods} to all 263 venues, clinical imaging journals ($n=131$)
carry reader assessment at nearly three times the rate of engineering and methods
venues ($n=78$), which in turn release code at nearly twice the clinical rate;
general neuroscience and neuroimaging venues ($n=37$) sit between. Supplementary
Table~S12 gives the full cross-tabulation. The rule leaves $17$ of the $263$ journal
titles unmatched, and those are reported as a fourth \emph{unclassified} group rather than
forced into one of the three; the four groups between them account for all $57$ reader
studies and all $97$ code releases. The tendency for the two practices to fall in different studies rather
than the same ones gives a Fisher exact odds ratio of $0.43$ ($p=0.013$). The estimate is
insensitive to the one labelling choice that could move it: taking the reader subset as the
61 studies originally labelled at extraction rather than the 57 verified at appraisal gives
an odds ratio of $0.43$ and $p=0.010$. We nonetheless report the association descriptively
rather than as a hypothesis test, because the venue grouping it rests on is a post-hoc
keyword rule that was not part of the fixed protocol, and a $p$-value computed on a
partition chosen after seeing the data does not carry its nominal meaning. The direction and
magnitude are what we rely on. The magnitude is the part that does not depend on the
choice: computational verifiability and diagnostic verifiability are largely held by
different communities, only one study in twenty holds both, and a safety argument
requires both.

\begin{figure*}[!tbp]
\centering
\includegraphics[width=\linewidth]{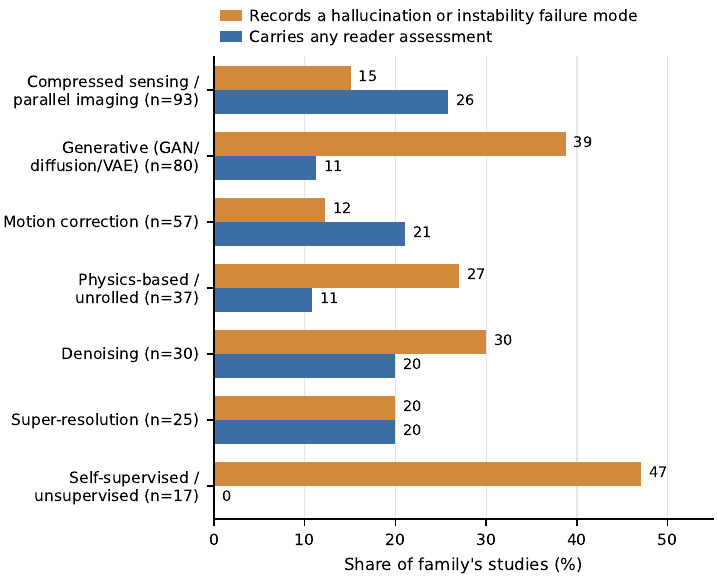}
\caption{Reader assessment against hallucination awareness, by method family
(recorded extraction labels; lower bounds). Bars are the share of each family's
studies that record a hallucination or instability failure mode (orange) and the
share that carry any reader assessment (blue), using the verified reader subset ($n=57$). Bar labels are rounded to the nearest integer; the one-decimal values quoted in the text are the same quantities (for example $15$ against $15.1\%$, $39$ against $38.8\%$). Generated from the
released per-study file (\texttt{included\_characteristics.csv}) by
\texttt{5\_Code/make\_figures.py}; every value is recomputable from the deposit.}
\label{fig:family_evaluation}
\end{figure*}

\emph{Awareness of a failure mode does not produce the evaluation that would detect it.}
Generative studies record hallucination or fidelity instability in $31/80$ ($38.8\%$)
of cases, the highest share among the four largest method families, against $15.1\%$ for
compressed-sensing and parallel-imaging work and $12.3\%$ for motion correction. Yet at
$9/80$ ($11.3\%$) they are also among the two least reader-evaluated of those families,
while the small self-supervised group records the corpus's highest hallucination share
($8/17$, $47.1\%$) and no reader assessment at all ($0/17$); per-family shares are in
Table~\ref{tab:practice_trends} and Fig.~\ref{fig:family_evaluation}. The sub-fields that most clearly recognise
that their models can synthesise structure are the least likely to run the test capable
of detecting it.

\begin{figure*}[!tbp]
\centering
\includegraphics[width=\linewidth]{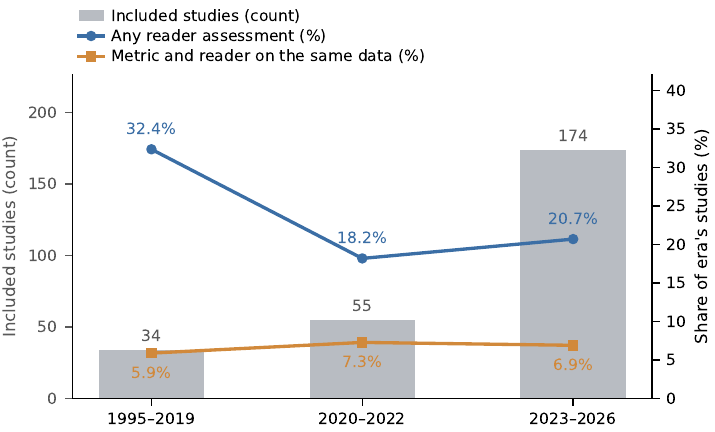}
\caption{Evaluation practice across publication eras. Grey bars give the number of
included studies per era (34, 55, 174); lines give the share with any reader
assessment and the share co-reporting a fidelity metric and a reader assessment on
the same data, using the verified reader subset ($n=57$). What recovery there is in
2023--2026 comes from subscription-walled clinical journals ($52.0\%$ reader assessment
in that stratum against $15.4\%$ in the open remainder; Section~\ref{sec:threats}).
Generated from the released per-study file by \texttt{5\_Code/make\_figures.py}; counts
are lower bounds from recorded extraction labels.}
\label{fig:practice_over_time}
\end{figure*}

\emph{Evaluation is not broadening as the field grows.} Reader-assessment share fell
from $32.4\%$ (1995--2019) to $18.2\%$ (2020--2022), recovering only partially to
$20.7\%$ (2023--2026) while the corpus grew fivefold
(Table~\ref{tab:practice_trends} and Fig.~\ref{fig:practice_over_time}). What recovery
there is comes entirely from the subscription-walled clinical literature: among
2023--2026 studies obtainable only through title-level library subscriptions, reader
assessment runs at $52.0\%$ (13 of 25), against $15.4\%$ (23 of 149) in the openly
accessible remainder of that era. Reader evidence, where it exists, sits
disproportionately behind paywalls, and a review that cannot retrieve those texts will
understate it; Section~\ref{sec:threats} takes the direction of that bias seriously
rather than dismissing it. Among the 144 studies recording any evaluation paradigm,
four in five record exactly one, and the share recording two or more fell across the
same periods (Supplementary Table~S12). Of the 84 studies recording a pixel-wise
metric, 60 ($71.4\%$) record no reader, uncertainty or perceptual measure beside it.

\begin{table*}[!tbp]
\centering
\caption{Evaluation practice across publication eras and method families (recorded labels; lower bounds). Reader-assessment share fell as the corpus grew and recovers only partially in the latest era, and that recovery is carried by subscription-walled clinical journals. Among the four largest method families the generative group, the one most associated with hallucination, is one of the two least reader-evaluated and is indistinguishable from physics-based unrolled work; the small self-supervised family ($n=17$) records no reader assessment at all. Reader shares use the verified reader subset ($n=57$; Section~\ref{sec:methods}), so counts sum to 57 across the era rows and are multi-label across the family rows.}
\label{tab:practice_trends}
\footnotesize
\setlength{\tabcolsep}{4pt}
\renewcommand{\arraystretch}{1.12}
\begin{tabular}{l r r r r}
\hline
\textbf{Era} & \textbf{n} & \textbf{Reader} & \textbf{Uncert.} & \textbf{Co-report} \\
\hline
1995--2019 & 34 & $32.4\%$ (11/34) & $5.9\%$ & $5.9\%$ \\
2020--2022 & 55 & $18.2\%$ (10/55) & $5.5\%$ & $7.3\%$ \\
2023--2026 & 174 & $20.7\%$ (36/174) & $7.5\%$ & $6.9\%$ \\
\hline
\multicolumn{5}{c}{} \\[-6pt]
\multicolumn{5}{l}{\textbf{Method family (multi-label)}} \\
\hline
\multicolumn{2}{l}{\textbf{Family}} & \textbf{n} & \multicolumn{2}{r}{\textbf{Reader share}} \\
\hline
\multicolumn{2}{l}{Compressed sensing / parallel imaging} & 93 & \multicolumn{2}{r}{$25.8\%$ (24/93)} \\
\multicolumn{2}{l}{Motion correction} & 57 & \multicolumn{2}{r}{$21.1\%$ (12/57)} \\
\multicolumn{2}{l}{Super-resolution} & 25 & \multicolumn{2}{r}{$20.0\%$ (5/25)} \\
\multicolumn{2}{l}{Denoising} & 30 & \multicolumn{2}{r}{$20.0\%$ (6/30)} \\
\multicolumn{2}{l}{Generative (GAN/diffusion/VAE)} & 80 & \multicolumn{2}{r}{$11.3\%$ (9/80)} \\
\multicolumn{2}{l}{Physics-based / unrolled} & 37 & \multicolumn{2}{r}{$10.8\%$ (4/37)} \\
\multicolumn{2}{l}{Self-supervised / unsupervised} & 17 & \multicolumn{2}{r}{$0.0\%$ (0/17)} \\
\hline
\end{tabular}
\end{table*}

\emph{Readers are not displaced away from the sequences where the stakes are highest.}
One expectation is not borne out. The comparison must be made on disjoint strata to mean
anything, because pulse sequence is a multi-label field and the naive contrast
double-counts the $107$ studies that acquire both an acute-pathology and a structural
sequence (Supplementary Table~S11 gives the correction). That the two disjoint strata below sum to $23+84=107$, the same number, is a third coincidence of unrelated tallies rather than a transcription error, verified against the released per-study file as the two flagged in Section~\ref{sec:requirements} are. Restricted to studies acquiring
one group and not the other, reader assessment runs at $5/23$ ($21.7\%$) for DWI or
FLAIR without T1w or T2w against $10/84$ ($11.9\%$) for T1w or T2w without DWI or FLAIR, representing a difference in the direction one would hope for, resting on 23 studies in the
smaller stratum and not approaching significance (Fisher exact odds ratio $2.06$,
$p=0.31$). We report it as descriptive and draw only the negative conclusion it
supports: there is no evidence that reader assessment is \emph{displaced away} from the
sequences on which acute pathology is read. The deficit is one of scarcity and of
distribution across research communities, not of misdirection within the clinical
literature.

\subsection{Methodological quality and risk of bias}
Across all 57 QUADAS-2 appraised diagnostic and reader studies, methodological quality was highest for Flow and Timing ($45/57$ low risk of bias, $78.9\%$), reflecting the standard retrospective undersampling paradigm where reconstruction and reference share the same acquisition timepoint without temporal degradation. Reference Standard and Index Test both sit at $26/57$ low risk ($45.6\%$), each now below half, with high ratings arising where radiologists scored accelerated images with reference ground truth visible or where lesion truth was defined by the comparator images themselves (incorporation bias). Patient Selection remains the weakest domain ($15/57$ low risk, $26.3\%$; $19/57$ unclear, $33.3\%$; $23/57$ high risk, $40.4\%$), as cohorts were predominantly retrospective, single-centre, and pathology-enriched without explicit prospective enrolment rules. Applicability concerns follow a similar distribution, with low concerns in Patient Selection for $36/57$ ($63.2\%$), Index Test for $30/57$ ($52.6\%$), and Reference Standard for $39/57$ ($68.4\%$).

The two rounds do not place the weaknesses in quite the same order, and the pooled shares above differ from the first round's in every one of the seven domains. Low-risk shares fell for Index Test ($53.8\%$ to $45.6\%$), applicability of the Index Test ($59.0\%$ to $52.6\%$), Flow and Timing ($82.1\%$ to $78.9\%$), Reference Standard ($46.2\%$ to $45.6\%$) and applicability of Patient Selection ($64.1\%$ to $63.2\%$); they rose for Patient Selection risk of bias ($23.1\%$ to $26.3\%$) and applicability of the Reference Standard ($59.0\%$ to $68.4\%$). The two largest movements concern the index test and the applicability of the reference standard, and the first of these runs against the appraised studies. The second-round studies are therefore not a favourable addition to the appraisal, which is why the pooled figures are the ones reported here. The first-round shares quoted here already incorporate the five studies re-appraised on recovered full text.

Corpus-wide reporting transparency evaluated against STARD principles revealed substantial omissions across the broader 263 studies: $109/263$ ($41.4\%$) provided no identifiable reference standard ($47$ stating none at all, Table~\ref{tab:quality_appraisal}, and a further $62$ naming a ground truth they left unspecified), scanner field strength was unreported in $54/263$ ($20.5\%$), and $166/263$ ($63.1\%$) did not make code openly available. Overall, $220/263$ studies ($83.7\%$) omitted at least one key methodological indicator (field strength, reference standard, or open code). Two directions must be kept apart here, and conflating them has caused confusion elsewhere in this literature. With respect to the \emph{reconstruction methods}, this reporting and selection pattern is anti-conservative: retrospective single-centre evaluation on curated data flatters reported performance relative to what prospective clinical deployment would show. With respect to \emph{this review's own conclusion}, the same pattern is conservative: if published performance is flattered and failure is under-reported, then the scarcity of safety-relevant evaluation we document is an understatement, and the true gap is at least as large as the one measured. The safety argument turns on the second direction and is not weakened by the first. Detailed domain-level judgments and supporting quotations are available in the open project repository.

\subsection{Requirements for safety-oriented evaluation}
\label{sec:requirements}
Read together, the synthesis implies five requirements that any evaluation intended to
establish reconstruction safety must meet, each stated with the corpus evidence that
motivates it and none met by more than a small minority of included studies.
\textbf{R1, focal task-based measurement}: every global fidelity score must be
accompanied by a region-of-interest or task-based observer measure, because spatial
averaging is the first mechanism by which lesion-scale error disappears ($60/84$, or
$71.4\%$, of pixel-metric studies record nothing beside the metric; this equals the
artifact share of the $R$-reporting subset in Table~\ref{tab:acceleration_spectrum}
($45/63$) to the decimal, which is a coincidence of two unrelated tallies rather than a
transcription error, verified against the released per-study file). \textbf{R2, paired
metric--reader reporting}: metric validity claims require a reader comparison on the
same data ($6.8\%$ of studies provide one). \textbf{R3, reference-independent validity}:
evaluation must remain interpretable when no registered artifact-free reference exists
($26.2\%$ rest solely on a fully sampled acquisition). \textbf{R4, robustness
under acquisition shift as a primary outcome}, reported as a result rather than a
caveat; robustness or distribution shift is recorded as a failure-mode concern in
$76/263$ studies ($28.9\%$), almost always in the discussion rather than as a
pre-specified endpoint.
\textbf{R5, hallucination-specific detection}, operating on the image side, since
measurement-space residual checks are provably blind to null-space edits. To these the
analysis in Section~\ref{res:gap_structure} adds a precondition: benchmarks containing
the pathology whose erasure is at issue, without which R1, R2 and R5 cannot be
exercised at all.

\section{Discussion}
\label{sec:discussion}

Three conclusions follow. First, the comparison on which the field's central proxy
rests is almost never made: no included study reports a rank correlation between a
pixel-wise fidelity metric and radiologist diagnostic assessment, and only 18 of 263
record the two on the same data at all. Where readers do appear, what is measured is
mostly agreement between readers, and it is weak and condition-dependent. The finding
therefore rests more on the absence of measurement than on the measurements that exist. Second, the reason is
structural, through the four mechanisms of Section~\ref{sec:results}: background
voxels average focal error away, the loss functions are sign-symmetric where diagnosis
is not, adversarial terms pay for invented texture, and every one of these metrics
assumes a registered reference that motion destroys. The mechanisms are mathematical
properties of the metrics themselves; how often each matters clinically is the part
the thin reader evidence leaves less certain. Third, no evaluation paradigm now in use
settles the safety question by itself, a conclusion resting on the
paradigm-by-paradigm synthesis of Section~\ref{sec:results}.

The analysis of where the gap comes from, which was not part of the fixed protocol and
is reported as exploratory throughout, changes what would follow from these findings if
it holds. The shortfall does not look like a uniform oversight; it co-occurs with two
structural features. Evaluation practice is divided between a community that releases code and a
community that runs readers, with only $4.9\%$ of studies doing both (an association
we interpret descriptively in Section~\ref{res:gap_structure}, its venue partition being
post-hoc), and the benchmark datasets named in this corpus
contain no acute pathology, so the failure mode of greatest clinical consequence is
not merely under-measured but unmeasurable on the datasets in common use. If that
association is causal, which this design cannot establish, recommendations addressed
to individual authors will not close the gap; what would be required is benchmark
construction and reporting conventions that force the two halves of the evidence
together.

This matters because accelerated reconstruction has left the research setting.
Deep-learning acceleration, denoising and image enhancement ship inside commercial MRI
products, several of them cleared as software as a medical device: one such tool, an
FDA-cleared vendor-agnostic DICOM-domain enhancement product, is evaluated in this
corpus by a prospective multicentre multireader trial \citep{bash2021deep}, which is
also a reminder that a cleared product can be validated on one sequence family and
deployed across others. Published guidance in both jurisdictions increasingly asks
manufacturers to demonstrate performance for the device's stated intended use and under
conditions representative of deployment rather than on curated development data: the
FDA, Health Canada and MHRA guiding principles on good machine learning practice ask
that datasets be representative of the intended patient population and that testing be
independent of training \citep{fda2021gmlp}, the EU AI Act imposes accuracy and
robustness obligations on high-risk AI systems \citep{euaiact2024}, and the Medical
Device Regulation requires clinical evidence appropriate to the device's intended
purpose \citep{eumdr2017}. Our reading of that guidance is that aggregate pixel-wise
fidelity cannot support a claim about small-lesion detection, because
\eqref{eq:dpsnr_identity} shows the aggregate does not move when the small lesion is
removed. That reading is ours: no regulator has reviewed or endorsed it, and none of the
three instruments names an image-quality metric.

A practical obstacle compounds it, and it is the reason the requirements in
Section~\ref{sec:requirements} are addressed to infrastructure rather than to authors.
Commercial inline pipelines export DICOM images after undisclosed normalisation, coil
combination and vendor post-processing, and the raw complex $k$-space typically never
leaves the scanner. Every method in Table~\ref{tab:rq3_metrics} that could detect a
prior-induced edit needs something the exported DICOM does not carry: the null-space
projection needs the forward operator and the coil sensitivities, a counterfactual audit
needs the ability to re-run the acquisition under a controlled change, and even a
measurement-space residual needs $Y$. Independent third-party auditing of a deployed
reconstruction is therefore not merely difficult but structurally impossible without an
open raw-data interface, and no amount of reporting discipline by authors of research
papers can supply one.

Three things follow that are worth stating as claims rather than as recommendations.

\emph{The cheapest useful change costs no new data.} A reconstruction paper that claims
its method preserves small lesions can state, from \eqref{eq:dpsnr_identity} and its own
brain volume and assumed local error ratio, the change in global PSNR that erasing the
smallest lesion it claims to preserve would produce. For a lacunar infarct that is about
$0.03$ dB, which is smaller than the difference the paper is likely to be reporting
between its method and its baseline. This is arithmetic, not a new experiment, and it
converts an implicit claim into an explicit one at the point where the claim is made.
Nothing in this review suggests such a line would be difficult; the finding is that
nobody currently reports it.

\emph{The two-community split is self-reinforcing, which is why exhortation has not
worked.} Each community's evaluation is close to the best it can do with the data it can
obtain. Engineering venues evaluate on the datasets that carry raw $k$-space, and those
contain healthy volunteers, chronic cohorts and tumours rather than acute pathology, so
a reader study run on them would not test lesion preservation even if it were run.
Clinical venues have the pathology, the readers and the acute presentations, and no
access to raw $k$-space or to the vendor's reconstruction, so a reproducibility claim is
not available to them. The $4.9\%$ of studies doing both are not more diligent than the
rest; they are the studies that happened to sit where both kinds of access coincide. An
instruction to individual authors to run more reader studies asks most of them to do
something the data available to them makes uninformative.

\emph{The missing object is a dataset, and its requirements are now specifiable.} The
public resources discussed in this review supply each necessary component separately, and
the distinction between what the corpus uses and what merely exists is part of the point:
raw multi-coil $k$-space is available in three of the eleven datasets the included studies
actually name (fastMRI, Calgary--Campinas, M4Raw); acute cerebrovascular pathology with
expert segmentation exists in ISLES and ATLAS, which no included study uses because they
are distributed as processed images; and reader-level pathology annotation on brain MRI
exists in fastMRI+, which no included study uses either. None combines them, and the combination is what
every requirement in Section~\ref{sec:requirements} presupposes: raw $k$-space so the
forward operator is known, acute lesions confirmed against a standard independent of the
images being evaluated, lesion-level annotation so a focal figure of merit can be
computed, and a licence permitting redistribution so the benchmark can be used by people
who did not collect it. That object does not exist, and until it does the safety
question is not merely unanswered but unaskable on public data.

\subsection{Priorities for future research}

Three of the five requirements in Section~\ref{sec:requirements} translate directly into
research priorities, and we do not restate them here: paired metric--reader reporting
(R2), robustness under acquisition shift reported as a primary endpoint rather than a
caveat (R4), and benchmarks carrying the pathology whose erasure is at issue, which is
the precondition on which R1, R2 and R5 all depend. Calibrated voxel-level uncertainty
that tracks local reconstruction error \citep{khawaled2024npbrec} belongs with R4 and is
the form in which R4 is most likely to be actionable for generative and self-supervised
architectures.

A fourth priority does not follow from anything this review extracted, and we separate
it for that reason. Many current pipelines operate slice by slice and introduce
inter-slice discontinuities on isotropic 3D acquisitions, so reconstruction and auditing
frameworks should model 3D anatomical coherence directly \citep{wei2026latent} while
meeting scanner-side latency constraints. Slice-wise versus volumetric processing was
not an extraction field in this review. We state the priority on mechanistic grounds and
on the cited work, not on a corpus count.

\subsection{Threats to validity}
\label{sec:threats}

The strongest conclusions rest on a small subset of reader-based studies, and we state
that prominently rather than in passing: 57 of 263 studies carry any reader assessment
and 18 pair it with a metric. Four threat classes bound what follows from that base.

\emph{How to read the limitations below.} They are not all bounded in the same way, and
treating them as one undifferentiated list of caveats would misrepresent what this review
knows about itself. Three kinds appear. Some are \emph{bounded by released evidence}: the
non-retrieval analysis, the paywall stratification and the appraisal-agreement
decomposition are each quantified from files a reader can recompute. Some are
\emph{bounded by argument}: the direction of publication bias and of the reporting
asymmetry is reasoned rather than measured, and we say which way each points. One is
simply \emph{unbounded}, and it is the sharpest: extraction has no reliability estimate.

Three properties of the extraction layer follow from that and should be weighed together.
First, its depth is deliberately shallow. Method, failure-mode and evaluation categories
were assigned from title, abstract and controlled vocabulary, so the prevalence counts are
floors on what studies \emph{report at that level} rather than measurements of what they
did. Second, the shallowness is visible in the counts themselves: only 84 of 263 studies
are recorded as reporting a pixel-wise fidelity metric, and 39 of the 57 verified reader
studies are recorded as reporting none, which for this literature is implausible as a
description of practice and entirely plausible as a description of abstracts. A further
119 of 263 studies ($45.2\%$) carry no evaluation-paradigm label at all, which is why we
report both the $/263$ and the $/144$ denominators wherever the distinction matters. Third,
the headline $18/263$ mixes two depths: the reader label was verified against full text and
corrected at appraisal, while the metric label was not. That asymmetry biases the
intersection downward by an unknown factor, and it is the single respect in which our
headline figure is least defensible as a point estimate, though the direction is
conservative, since correcting it could only widen the measured gap.

What we offer in place of a reliability coefficient is auditability, and the substitution is
deliberate rather than a shortfall we failed to notice. Every extracted value is released
with the verbatim quotation that supports it, so a reader can check whether an assignment is
\emph{correct}, which is a stronger and more useful guarantee than a coefficient showing
that two people were \emph{consistent}. It is not a substitute for reliability in the
statistical sense, and we do not claim it is: consistency and correctness bound different
failure modes, and only the first is estimable from our records. A duplicate-extraction
exercise on a random subsample, reported with a $\kappa$, is the obvious next step and we
name it as an outstanding task rather than a completed one.

\emph{Construct validity.} We searched with broad, high-sensitivity terms for
architectures, artifacts and evaluation paradigms, and the corpus spans the
reconstruction-safety construct, from pixel-wise metrics through perceptual losses to
reader studies. What the corpus cannot guarantee is that a metric means the same thing
wherever it appears: studies differ in what they treat as the quantity of interest, so
a rank correlation with reader scores in one paper is not straightforwardly the same
measurement as in another, and grouping such values under one heading is an
interpretation we make, not a property of the data. The venue classification
underlying Section~\ref{res:gap_structure} is likewise a post-hoc keyword assignment,
released with the data rather than extracted from the studies.

\emph{Internal validity.} A protocol fixed before screening began, a documented search
across seven databases, and independent screening and appraisal by three reviewers provide
robust protection, and every extracted value is released with the verbatim snippet supporting
it. Five specific threats bound these findings.
First, $81$ of the $1{,}149$ candidate reports remain without obtainable text through
structural licensing barriers. Screening the nearest obtainable stratum (the $341$
reports recovered only through title-level subscriptions) found $86.2\%$ ineligible on
the same criteria, so at that rate the $81$ would contribute on the order of $11$ studies
rather than $81$. That bounds the missing evidence without establishing that the bias is
nil. Non-retrieval is demonstrably \emph{not} random with respect to the variable this
review measures: within the same stratum, 2023--2026 studies carry reader assessment at
$52.0\%$ against $15.4\%$ in the openly accessible literature of that era. Both findings
point the same way. Reader evidence is over-represented behind paywalls, so a review that
cannot retrieve those texts understates how much reader assessment the field performs, and
our headline scarcity figures are conservative bounds rather than point estimates.
Second, four included studies were assessed on abstract and controlled vocabulary alone and one on partial text; a sensitivity analysis recomputing all headline metrics on the $258$ studies verified against complete full text leaves every finding directionally stable (e.g., metric--reader co-reporting remains $7.0\%$ vs $6.8\%$).
Third, every QUADAS-2 tally in this article rests on the whole verified reader subset ($n=57$, $399$ domain judgements), so none is computed on a partial base. The second round moved all seven domain distributions, and the movements that most concern the index test run against the appraised studies: low risk of bias for the Index Test falls from $53.8\%$ to $45.6\%$, and high applicability concern for the Index Test rises from $33.3\%$ to $40.4\%$ (Section~\ref{sec:results}). Four studies remain abstract-only and are appraised on that basis.
Fourth, reliability is measured for screening and for appraisal but not for extraction,
and extraction is the sharpest methodological limitation of the review. The screening and
appraisal coefficients, and the decomposition that must be read with the appraisal figure,
are given in Section~\ref{sec:methods} and not restated here; the consequence for this
section is that $126$ of the $399$ QUADAS-2 judgements were reached at a level of agreement
a single rater could not be relied on to reproduce, so the domain shares in
Table~\ref{tab:quality_appraisal} carry that uncertainty. It does not propagate to the
prevalence findings, which rest on extraction labels rather than QUADAS-2 ratings, but it
bounds how firmly the risk-of-bias picture can be stated. Neither screening nor appraisal
agreement transfers to extraction. The duplicate-screening sample is also smaller than the
$300$--$400$ records a referee might reasonably ask for, and the interval reflects that.
Fifth, four studies recorded at extraction as observer or reader designs were found at
appraisal to contain no human reader and were reclassified as algorithmic, moving the
reader subset from $61$ to $57$ and the algorithmic set from $202$ to $206$. The effect on
the headline numbers is small and in the direction of a larger gap: metric--reader
co-reporting moves from $7.6\%$ to $6.8\%$, the 2023--2026 reader share from $23.0\%$ to
$20.7\%$, the code-and-reader intersection from $5.3\%$ to $4.9\%$, and the Fisher exact
$p$ from $0.010$ to $0.013$ at an odds ratio that moves only from $0.43$ to $0.43$. Those
four corrections are also the source of the $4/61$ ($6.6\%$) error rate quoted above, and
the design field is the only one that received three-reviewer full-text verification: a
comparable rate elsewhere would not overturn any finding here, since the headline shares
are separated from their comparators by more than that margin, but it is a source of
variability measured in one place and assumed in the rest.

\emph{External validity.} Among the $209$ studies stating a field strength the corpus
covers 1.5, 3 and 7 Tesla ($21$, $153$ and $35$ respectively), together with several
vendors' coil configurations and the major pulse sequences; the remaining $54$ ($20.5\%$)
state none, and nothing about field-strength coverage should be read into them. Tasks,
protocols, architectures, datasets and metrics otherwise vary substantially, which is why
we report the direction and consistency of findings rather than their magnitudes. Many
algorithmic studies evaluate on retrospectively undersampled single-centre data, which
flatters performance relative to prospective multi-centre use; where the public data were
themselves pre-processed before undersampling, that inflation is larger again
\citep{shimron2022implicit}. Prospective multicentre reader evidence is thin, and the
examples this review leans on each carry a boundary that must travel with them: the
emergency-department comparison is single-centre and its one discrepant finding is a single
case \citep{lang2024ultrafast}; the widest-ranging acceleration reader study is
retrospective and single-centre \citep{radmanesh2022exploring}; and the largest prospective
multicentre multireader trial evaluated 3D T1-weighted acquisitions only, with an
image-domain DICOM enhancement tool rather than reconstruction from undersampled
$k$-space, reported improved rather than degraded lesion conspicuity, and has several
authors affiliated with the vendor of the evaluated tool \citep{bash2021deep}; we cite
it for what it does establish, that such tools are cleared and deployed. Finally, the
benchmark concentration of Section~\ref{res:gap_structure} (eleven public datasets named
across the corpus, not one curated for acute ischaemic stroke or intracranial haemorrhage
) caps how far any leaderboard result can be carried into emergency imaging.

\emph{Conclusion validity.} Because pixel-wise metrics agree only loosely with
radiologist judgement (RQ1), a study resting on PSNR and SSIM alone supports a safety
claim less well than one with readers or uncertainty-aware observers, and we weight
the evidence accordingly. Publication bias favouring successful reconstructions is a
material limitation and works in the same direction, making the reported prevalence of
failure evidence an understatement: a literature with little incentive to report
failure is exactly the corpus a review about failure must draw on, and the absence of
a common effect measure rules out funnel-plot and small-study asymmetry tests. We
applied no certainty framework such as GRADE, since the review produces no pooled
effect estimates for such a framework to rate. Prevalence counts derive from recorded
extraction labels and understate true practice, and no range we quote should be read
as an interval estimate rather than as values from named studies.

\section{Conclusion}
\label{sec:conclusion}

We reviewed 263 primary studies on the safety, artifacts and fidelity of deep
learning-based brain MRI reconstruction. Wherever metrics and readers have been compared, agreement between conventional fidelity scores and radiologist judgement is weak, and at most $18$ of the $263$ studies record the paired metric--reader evaluation capable of exposing that divergence. Every prevalence figure here is a floor derived from screening-level labels rather than a measurement of practice, and although screening agreement was high, the extraction labels those floors rest on carry no reliability estimate of their own. The weakness is structural rather than incidental: the four mechanisms identified under RQ2 explain why, and the safety question is not one that any current evaluation paradigm answers alone.

The contribution we would emphasise is the account of why the gap persists, and we
offer it as a hypothesis these data are consistent with rather than as an established
structural fact: the venue grouping it rests on is post-hoc, so although the association
between code release and reader assessment is strong and would be conventionally
significant on a pre-specified partition, we do not read it as a test
(Section~\ref{res:gap_structure}). What does not depend on that classification is the
magnitude. Only $4.9\%$ of studies both release code and run readers, so computational
verifiability and diagnostic verifiability are, as recorded, held by different studies
roughly nineteen times in twenty; and no dataset named in this corpus is curated for
acute stroke or intracranial haemorrhage. A reconstruction model can
therefore satisfy every metric and every public benchmark in current use while the
question of whether it preserves a small acute lesion has never been asked. Closing that
gap requires benchmarks that contain the pathology at issue and reporting conventions
under which no pixel-wise metric appears alone, and until both exist, claims that
accelerated reconstruction is diagnostically safe will continue to rest on measurements
that cannot see the failures that matter.

One change is available immediately and needs no new data. Equation~\eqref{eq:dpsnr_identity}
converts any lesion volume into the PSNR movement its erasure would cause, so an author
reporting a reconstruction at a stated acceleration can state, alongside the global
score, the smallest focal change that score is capable of registering at their image
size. For a $100\text{ mm}^3$ infarct that figure is around $0.03\text{ dB}$, which is
below the precision at which PSNR is conventionally reported. Publishing that bound next
to the metric costs one line, requires no reader and no new acquisition, and makes
explicit what the metric is silent about. It is the cheapest useful step towards
evaluation that could detect the failure this review is concerned with.

\clearpage

\section*{Funding Statement}
This research was supported by institutional research computing infrastructure and scholarship funding from RMIT University. The funder had no role in the design of the review, the search strategy, the eligibility criteria, the screening or appraisal of studies, the synthesis or interpretation of findings, the writing of this report, or the decision to submit it for publication. No external, commercial, or industry funding was received, and no vendor of MRI hardware or reconstruction software contributed financial or non-financial support to this work.

\section*{Ethics statement}
This review synthesises secondary data from published, de-identified study reports. It
involved no human participants, no animal experimentation and no collection of private
health data, so institutional review board approval and patient informed consent were
not required.

\section*{Data and code availability}
The review protocol, executed search strategy, screening criteria, every title/abstract
and full-text decision with its recorded reason, the completed PRISMA 2020 checklist,
the per-study extraction and QUADAS-2 worksheets for all 263 included studies, the
analysis code, and the duplicate-screening sample and per-reviewer appraisal worksheets
with their agreement computations (\texttt{kappa\_worksheet\_214\_HUMAN\_*.csv},
\texttt{kappa\_human\_results.csv}, \texttt{reviewer\_worksheet\_R*.csv} and
\texttt{quadas2\_three\_reviewer\_worksheet.csv}) are openly available on the Open Science Framework at
\url{https://doi.org/10.17605/OSF.IO/832Z6} and mirrored at
\url{https://github.com/dmai287/brain-mri-recon-safety-slr}, archived at Zenodo under
DOI \href{https://doi.org/10.5281/zenodo.22100024}{\nolinkurl{10.5281/zenodo.22100024}}
(release \texttt{v1.0.3}, commit \texttt{8c856ba}; concept DOI
\href{https://doi.org/10.5281/zenodo.21836966}{\nolinkurl{10.5281/zenodo.21836966}}). Every count
reported here is regenerable from \texttt{included\_characteristics.csv} in that
release, and the deposit encodes the same reader/algorithmic partition ($57$/$206$) that
this article reports, with the four reclassified studies flagged individually. The two
figures generated from that file and the $\Delta$PSNR identity figure are produced by
\texttt{5\_Code/make\_figures.py} and \texttt{5\_Code/make\_dpsnr\_figure.py}. The deposit is retrospective and is documented as such. Full-text PDFs of the
included studies are not redistributed, as publisher licence terms do not permit it; the
DOI list in the released bibliography (\texttt{included\_studies\_263.bib})
identifies the corpus in full.

\section*{CRediT authorship contribution statement}
\textbf{Dat Tat Mai:} Conceptualization, Methodology, Software, Investigation,
Data curation, Formal analysis, Visualization, Project administration,
Writing -- original draft, Writing -- review \& editing.
\textbf{Thai Viet Pham:} Investigation, Data curation, Validation, Resources,
Writing -- review \& editing.
\textbf{Thu Nguyen Thi Dang:} Investigation, Data curation, Validation,
Writing -- review \& editing.
\textbf{James Jin Kang:} Supervision, Validation, Writing -- review \& editing.

\section*{Declaration of competing interest}
The authors declare that they have no known competing financial interests or
personal relationships that could have appeared to influence the work reported
in this paper.

\section*{Declaration of generative AI and AI-assisted technologies in the manuscript preparation process}
During the preparation of this work the authors used generative AI technologies solely to
assist with language editing and proofreading of the manuscript text. After using these
tools, the authors reviewed and edited the content as needed and take full responsibility
for the content of the published article. All study screening, data extraction, and quality
appraisals reported in this review were conducted manually by the authors.

\section*{Appendix A. The included studies and how to trace a claim to them}
All $263$ included primary studies appear in the reference list, so the evidence base is
part of the article of record rather than of the supplement alone. Because the synthesis is
narrative and most claims are grouped rather than attached to a single study, the reference
list alone does not show which studies produced a given number. That mapping is carried by
\texttt{included\_characteristics.csv}, which holds one row per included study with its
extracted method family, evaluation paradigm, failure modes, sequences, field strength,
reference standard and code availability, each with the verbatim quotation supporting it,
so any count, share or crosstab in Section~\ref{sec:results} can be reproduced and the
contributing studies listed by filtering that file. Supplementary Tables~S1--S3 give the
same $263$ studies as a numbered index by publication year, and
\texttt{included\_characteristics\_supplement.csv} adds the per-study QUADAS-2 and
reproducibility ratings.

The reference list therefore contains two kinds of entry, and a reader needs a rule to tell
them apart. It prints $299$: the $263$ included studies, carried into the list under PRISMA
2020 item 17, and $36$ narrative-only works cited in argument. \textbf{An entry appearing in
the numbered index of Supplementary Tables~S1--S3 is a member of the reviewed corpus; an
entry that does not is a narrative citation.} Equivalently, and checkable without the
supplement, every narrative citation is cited at a specific point in the running text of
Sections~1--6, whereas corpus entries not discussed by name appear without an in-text
citation. The two lists were generated by independent mechanisms and agree exactly as sets.
We would be glad to move the corpus to a separately labelled supplementary reference list
if the editor prefers that production route.

\section*{Appendix B. Supplementary material}
Supplementary material associated with this article (Supplementary Tables~S1--S12 and
Supplementary Figure~S1) is provided as a separate document. Tables~S1--S3 list all
263 included primary studies with full bibliographic details; Tables~S4--S9 give the
comparison against related secondary studies, the search vocabulary and its derivation,
the extraction form, the thematic evidence architecture, the four metric-failure
mechanisms with their evidence sources, and the provenance of every released data file.
Tables~S10--S12 carry the supporting detail for Section~\ref{res:gap_structure}:
the annotation-layer derivation for fastMRI+ (S10), the disjoint-strata correction for
the sequence contrast (S11), and the venue cross-tabulation with evaluation-paradigm
multiplicity by era (S12).

\nocite{gardner1995detection,cao1995using,fadeev2005quality,tisdall2006using,miao2013a,li2015artifactual,ning2015sparse,elhabian2016compressive,lacerda2016diffusion,zahneisen2016reverse,esteban2017mriqc,yu2018a,gurbani2018a,fantini2018automatic,osadebey2018blind,risk2018impacts,tang2019accelerated,zhao2019applications,sujit2019automated,vranic2019compressed,glessgen2019evaluation,hagiwara2019improving,haskell2019network,bollmann2019sharqnet,elsaid2019superresolution,sommer2020correction,benjamin2020diagnostic,hampe2020investigating,vanderhasselt2020synthetic,zhang2021apiremc,ding2021acceleration,harper2021assessing,duong2021correcting,bash2021deep,sagawa2021deep,kwan2021deep,karimi2021deep,dieckmeyer2021effect,hu2021runup,muckley2021results,duffy2021retrospective,kim2021simultaneous,gao2021xqsm,zhang2022a,zibetti2022alternating,conklin2022clinical,sheng2022deep,huang2022deep,begnoche2022epi,radmanesh2022exploring,sannananja2022imagequality,yasaka2022impact,muro2022improvement,pirkl2022learning,yang2022modelbased,ma2022mitigating,yoshida2022motion,g2022quantitative,srikant2022quantitative,sui2022scanspecific,hu2022spatiotemporal,pawar2022suppressing,qiao2022unsupervised,gngr2023adaptive,lang2023clinical,johannes2023comparison,chen2023ddcisenet,pizarro2023deep,hossbach2023deep,li2023diffusion,shan2023distortioncorrected,thaler2023effect,usui2023evaluation,ciceri2023geometric,ramosllordn2023highfidelity,ji2023highly,simk2023improving,jiang2023multimodal,tabari2023optimized,nivetha2023sadir,yilmaz2023selfsupervised,wang2023spatialintensity,beljaards2024aibased,i2024accelerated,zhou2024adversarial,dolui2024automated,heo2024cest,karthik2024comprehensive,lee2024deep,uher2024deepflair,bian2024improving,yasaka2024iterative,trivedi2024mcddpm,ma2024mri,khawaled2024npbrec,kofler2024quantitative,gil2024quantitative,qiu2024selfcalibrated,olsson2024simulating,rizzuti2024towards,altmann2024ultrafast,jun2024zerodeepsub,feizollah2025d,won2025a,safari2025a,zhu2025a,nishioka2025accelerating,wen2025accelerating,marchetto2025agreement,liu2025application,bjrkeli2025artificial,angella2025assessing,nonninger2025assessment,hosseini2025autocornn,yarach2025blipup,you2025conditional,zhu2025cycleconditional,kharaji2025dantecaipi,angella2025dima,liebrand2025deep,giulio2025diffusion,cerjanic2025elastographic,yoo2025evaluation,tang2025incorporating,nimje2025insights,mercier2025intersectionbased,singh2025learningbased,raymond2025leveraging,taymourtash2025measuring,karuna2025modified,goffi2025multimetric,byanju2025myelin,karl2025nonreference,chen2025pwgan,li2025patientspecific,sepideh2025probabilistic,choi2025prospective,jochmann2025quantitative,wichtmann2025rapid,wang2025rapid,verdicchio2025reliability,jung2025reliability,safari2025resmocodiff,aali2025robust,dong2025romerepti,dabrowski2025sismik,huber2025shortterm,alalar2025sparsitydriven,wang2025spherical,dewey2025superresolution,liu2025timeefficient,bauman2025ultrahighresolution,muro2025using,wang2025variableflipangle,rauland2025white,huabing2026accelerating,xiong2026advancing,schauman2026an,thomas2026automatic,liu2026blockinterleaved,lee2026cartesian,okell2026combined,marchetto2026contrastoptimized,beljaards2026deepdisorder,yoshida2026deep,baz2026deep,s2026evaluating,ho2026evaluation,mello2026fast,li2026fewshot,b2026ganbased,liu2026highly,zhang2026iterative,wei2026latent,junhyeok2026lesionaware,abed2026mriqt,ortizgonzalez2026motiondps,shen2026realdeal,rassmann2026regression,f2026simulationbased,islam2026synpoc,sebastian2026ultrafast,kim2026vessel,ryu2019data,du2020brain,hales2020combined,kulkarni2021dictionary,tanno2021uncertainty,almasni2022stacked,kanemaru2022effect,vakli2023automatic,corderogrande2023fetal,zhou2023mitigating,lau2023pushing,ernst2023sinogram,yuan2024novel,wada2024deep,hewlett2024deep,xie2024based,moshe2024handling,desalegn2024hara,wu2025application,sakitis2025bayesian,brunatova2025denoising,lan2025diffusion,huijben2025enhancing,behrendt2025guided,park2025higher,meng2025motion,li2025quantitative,nagaraj2025slice,salem2026bald,starck2026diff,saeed2026improved,liang2026itermask,mirza2026learning,seonghyeon2026simulation,tessema2026technical,miao2010comprehensive,kayvanrad2016diagnostic,kim2017artificial,chow2017modifiedbrisque,gong2018deep,delattre2019compressed,zhang2019mri,chung2020restoration,zhang2019robust,liu2020motion,wang2020synthesize,meng2020accelerating,karimi2021accurate,garmaoehmichen2021evaluation,kumar2021learning,kim2022deep,chan2022signal,song2022jointly,wismuller2022selfsupervised,matsuo2023feasibility,yarach2024reconstruction,shan2024image,chen2025mri,ismail2024performance,wu2025d,choi2025accelerated,joshi2025arterial,liu2025deep,mohsen2025deep,xu2025deism,wang2025estimating,nagaraj2025evaluation,takada2025impact,huang2025improve,harada2025verification,milovic2024xsim,xia2026accurate,solak2026braingan,tabari2026clinical,jayadeepthi2026deep,verclytte2026deep,bathla2025deeplearning,cao2026loddpm,fujita2025motioninformed,lang2024ultrafast}

{\def\baselinestretch{1}\small
\bibliographystyle{IEEEtranN}
\bibliography{ref2}
}

\end{document}